\documentclass[12pt, draftclsnofoot, onecolumn]{IEEEtran}

\usepackage{cite}
\usepackage{url}
\usepackage[pdftex]{graphicx}
\usepackage{xcolor}
\usepackage{amsmath}
\usepackage{amsfonts}
\usepackage{amssymb}
\usepackage[caption=false,font=footnotesize]{subfig}
\usepackage{gensymb}
\usepackage{comment}
\newcounter{MYtempeqncnt}
\begin{document}

\title{Performance Analysis of Backscatter Communication Systems with Non-orthogonal Multiple Access in Nakagami Fading Channels}

 \author{Ahsan~Waleed~Nazar,
         Syed~Ali~Hassan,~\IEEEmembership{Senior~Member,~IEEE,}
         Haejoon~Jung,~\IEEEmembership{Member,~IEEE,}
         Aamir~Mahmood,~\IEEEmembership{Senior~Member,~IEEE,}
         and~Mikael~Gidlund,~\IEEEmembership{Senior~Member,~IEEE}
         
\thanks{A. W. Nazar and  S. A. Hassan are with the  School of Electrical Engineering and Computer  Science~(SEECS), National University of Sciences and Technology~(NUST), Islamabad, Pakistan. (e-mail: anazar.msee23mcs@students.mcs.edu.pk, ali.hassan@seecs.edu.pk).}
  
\thanks{H. Jung is with the Department of Information and Telecommunication Engineering, Incheon National University, Incheon  22012, Korea (e-mail: haejoonjung@inu.ac.kr).}
  
\thanks{A. Mahmood and M. Gidlund are with the Department of Information Systems and Technology, Mid Sweden University, Sweden (emails: aamir.mahmood@miun.se, mikael.gidlund@miun.se).}}

\maketitle
\begin{abstract}
Backscatter communication (BackCom) has been emerging as a prospective candidate in tackling lifetime management problems for massively deployed Internet-of-Things devices, which suffer from battery-related issues, i.e., replacements, charging, and recycling. This passive sensing approach allows a backscatter sensor node (BSN) to transmit information by reflecting the incident signal from a carrier emitter without initiating its transmission. To multiplex multiple BSNs, power-domain non-orthogonal multiple access (NOMA), which is a prime candidate for multiple access in beyond 5G systems, is fully exploited in this work. Recently, considerable attention has been devoted to the NOMA-aided BackCom networks in the context of outage probabilities and system throughput. However, the closed-form expressions of bit error rate (BER) for such a system have not been studied. In this paper, we present the design and analysis of a NOMA enhanced bistatic BackCom system for a battery-less smart communication paradigm. Specifically, we derive the closed-form BER expressions for a cluster of two devices in a bistatic BackCom system employing NOMA with imperfect successive interference cancellation under Nakagami-$m$ fading channel. The obtained expressions are utilized to evaluate the reflection coefficients of devices needed for the most favorable system performance. Our results also show that NOMA-BackCom achieves better data throughput compared to the orthogonal multiple access-time domain multiple access schemes (OMA-TDMA).

\end{abstract}

\begin{IEEEkeywords}
 IoT, NOMA, Nakagami-$m$ fading, backscatter communication, bit error rate.
\end{IEEEkeywords}

%
\IEEEpeerreviewmaketitle

\section{Introduction}

\textcolor{black}{Power-efficient wireless connectivity is at the heart of the massive proliferation of Internet-of-thing (IoT) networks for wide-scale and fine-grained sensing and control in various domains.  For instance, IoT networks of constrained battery-operated devices are attractive for applications in both the consumer and industrial domains, e.g., smart home/city/agriculture, e-health, building/industrial- automation, and vehicular/aerial networks \cite{dawy_toward_2017}. This pervasiveness and usability of IoT connectivity are equally reflected in currently connected IoT devices, which will further be reaching 24.5 billion devices by 2025 \cite{jejdling_ericsson_2020}. In this expected scenario, managing the network lifetime becomes critical, especially when the conventional battery-based solutions are not viable due to the high cost of battery replacements and recycling concerns. In particular, battery recharging/replacement is challenging in cases where most of the sensors are hidden (e.g., inside concrete walls) or deployed in an inaccessible environment (e.g., under cultivated land). As a result, various energy harvesting techniques are under investigation to overcome this challenge \cite{xu_internet_2014}.}

\textcolor{black}{Recently, for massive IoT networks, backscatter communication (BackCom) has emerged as one of the promising radio-frequency (RF) energy harvesting techniques, which enables communication without battery backup \cite{niyato_wireless_2017} \cite{Bletsas2018}.}
The low-complexity and low-power BackCom technique allows backscatter sensor nodes (BSNs) to communicate with a backscatter sensor reader (BSR) by passive reflection from a carrier emitter (CE) and modulation without requiring any active RF transmission component.
This backscatter is made possible because of the intentional impedance mismatch at the antenna input which results in different reflection coefficients \cite{lu_ambient_2018}. Data encoding for backscatter over an incident wave is carried out by varying the reflection coefficients at the antenna input side. This backscatter approach is vastly different from the generally applied wireless harvesting approach where nodes first harvest energy to perform active RF transmission. 
Thus, transmissions in a backscatter communication consume orders of magnitude less energy than a typical radio and the absence of an active RF component in BSNs enables simpler and low complexity circuits. Therefore, the BackCom approach is most suitable for IoT networks \cite{liu_next_2019}.

The backscatter technique implementation for passive IoTs has been limited due to its inherent short-range communication. Recently, the BackCom system has been suggested for overcoming this limitation using bistatic architectures \cite{Kimionis2014}, 
\cite{Alevizos2018}. In bisatic architecture, the CE and BSR are not co-located, therefore, it allows setting up a more flexible network topology which can also mitigate the near-far effect. 
Non-orthogonal multiple access (NOMA), because of its low latency and high spectral efficiency, is the ideal candidate to support a large number of IoT nodes in a BackCom uplink communication system \cite{ding_survey_2017}. Recently, NOMA-aided BackCom has proven to be a key technology for collating data from multiple BSNs \cite{Basnayake2020}, 
\cite{psomas_backscatter_2017}. In \cite{guo_design_2018}, the authors investigated the performance of a NOMA-enhanced BackCom system, and the significance of adopting NOMA with the BackCom system was demonstrated by analyzing the average number of successfully decoded bits. Similarly, the authors in \cite{zeb_noma_2019} and \cite{Le2019} evaluated the performance of NOMA-aided BackCom network in terms of outage probability and system throughput. The authors in \cite{Yang2019} and \cite{Liao2020} studied the problem of resource allocation in NOMA-enhanced BackCom networks. In \cite{Zhang2019}, the authors analyzed the outage probabilities and the ergodic rate for a symbiotic system that integrates cellular NOMA and ambient BackCom in an IoT network. The authors in \cite{Lyu2017} studied the optimal time allocation policies of a power station-powered BackCom system for a hybrid NOMA-TDMA scheme. Relay assisted BackCom system has been studied in \cite{Lyu2019}. In \cite{Farajzadeh2019}, the authors proposed a NOMA-BackCom application in unmanned aerial vehicle (UAV) based data collection with optimized UAV altitude and trajectory.

Based on the aforementioned literature survey, to the best of authors' knowledge, there is no reported work that carries out the BER performance analysis of uplink NOMA in a BackCom system, which is an essential component in system design and optimization. In this work, we consider a NOMA-enhanced bistatic backscatter uplink communication under the Nakagami-$m$ fading channel where one reader serves a cluster of randomly deployed BSNs. To maintain low NOMA decoding complexity, BSNs are usually divided into a cluster of two users. The reflection coefficients of the BSNs are set different from each other to make the wireless channel distinct to better exploit power-domain NOMA. 
The proposed scheme results in an increase in effective non-erroneous transmitted bits as compared to an OMA scheme. The main contributions of this paper are summarized as follows:

\begin{itemize}
  \item We propose a NOMA enhanced bistatic BackCom system for battery-less smart communication among BSNs, where the reflection coefficients of BSNs can be manipulated to achieve better system performance in terms of effective data bits. 
  \item The BER performance of a NOMA enhanced bistatic BackCom system, impaired by a Nakagami-$m$ fading channel, with imperfect SIC is considered, where approximate closed-form analytical BER expressions of binary phase-shift keying (BPSK) are derived for two BSNs. The derived BER expressions are verified by Monte Carlo simulations for various reflection coefficients.
  \item The probability density function (PDF) of the difference of two independent and non-identically distributed (i.n.i.d) Nakagami-$m$ distributions has not been derived in literature before. It is shown with the help of some statistical approaches such as the moment matching method that it follows a normal distribution and we derive the average BER expressions under such a scenario.
  \item The increase in effective non-erroneous transmitted bits over a large period in a NOMA-BackCom scheme is compared with an OMA scheme for a variety of  reflection coefficient conditions. It is shown that NOMA-BackCom outperforms the OMA-TDMA scheme by multiplexing the BSNs.
  \item Our results suggest that at relatively higher transmit SNRs, system performance can be greatly improved by setting suitable reflection coefficients rather than an increment in transmit power of the CE.
\end{itemize}

The rest of the paper is organized as follows. In Sec. II, the system model of the NOMA-aided BackCom system is briefly described. It is followed by the derivation of closed-form BER expressions of BPSK constellation for two BSNs in Sec. III. Numerical and simulation results are shown in Sec. IV. Finally, conclusions are presented in Sec. V.

\section{System Model}
\subsection{BackCom Model}
\begin{figure}[t]
  \centering
  \includegraphics[width= 0.9\columnwidth]{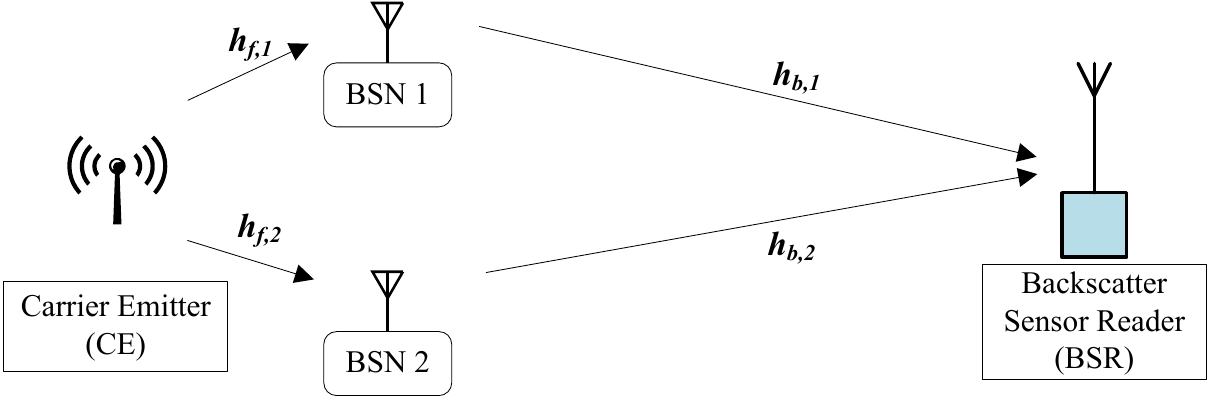}
  \caption{Illustration of uplink NOMA-aided BBCS.}
  \label{fig:System_model}
\end{figure}

We consider a NOMA-aided bistatic BackCom system (BBCS) consisting of a CE, multiple BSNs, and a BSR as shown in Fig. \ref{fig:System_model}. In practice, BSNs are usually multiplexed into different clusters of two or three users to maintain low decoding complexity and meet timing constraints. Moreover, CEs are placed closer to the BSNs in the field to mitigate the doubly near-far effect. The BSNs do not possess any active RF transmission source. The CE transmits a sinusoidal continuous wave (CW) carrier signal which is intelligently reflected by BSNs to aid in communication with BSR\@. The CW signal is transmitted by CEs all the time, whereas BSNs operate in two states, namely the active state and energy harvesting state.

In the \emph{active state}, each cluster of BSNs backscatters the CW signal to transmit its sensor data to the BSR\@. The BSR receives the signal and recovers the information from BSNs. BSNs are capable of reflecting the incident CW signal with altered phase and/or magnitude by terminating its antenna between two load impedances. This is typically carried out with the help of an RF transistor. Ideally, by switching the transistor on, the antenna is short-circuited and the incident wave is reflected with a phase change of 180\degree. Alternatively, by switching the transistor off, the antenna is open-circuited and the incident wave is reflected with no phase change (i.e., zero degree). By changing the value of load impedance from short and open-circuit conditions, the magnitude of the reflected wave can also be changed alongside phase.

In the \emph{energy harvesting state}, the BSNs do not reflect the incident CW signal, but only harvest the energy from it. The harvested energy is stored in a battery and is used to power the circuitry (including micro-controller and RF transistor) and support sensing operations. We assume that the cluster which is not in an active state is in an energy harvesting state or vice versa. Furthermore, the stored energy can be used to sustain long term operations of BSNs\@. 

\begin{figure}[ht]
  \centering
  \includegraphics[width=\columnwidth]{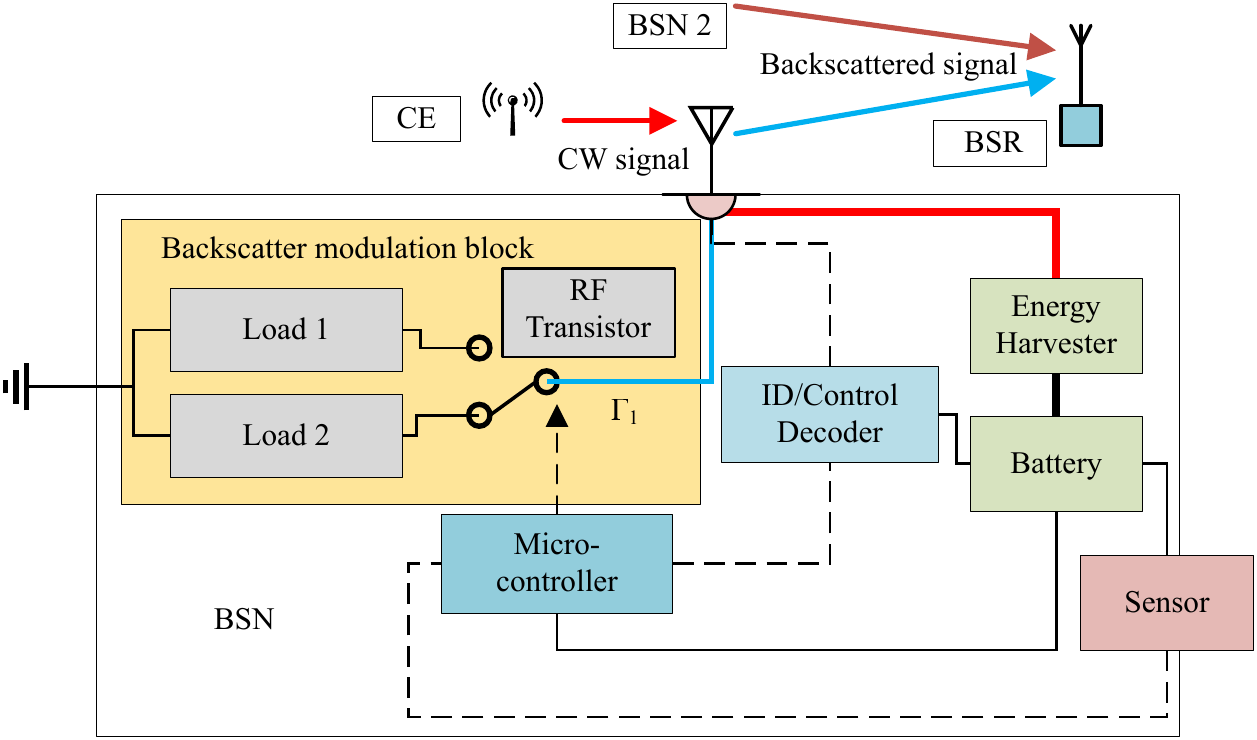}
  \caption{Illustration of the internal structure of a BSN.}
  \label{fig:tag}
\end{figure}

As the modulation of BSNs is achieved by switching the antenna loads between \(M\) load impedances, it corresponds to the power reflection coefficients, \(\Gamma_n\), \(n =\) \{0\,,\,.\,.\,.\,, \( M-1\)\}. Keeping in view the complexity and energy constraints of the low-power BSNs, we consider BPSK modulation in this work. Therefore, we only consider two values of power reflection coefficient \{\(\Gamma_0\), \(\Gamma_1\)\} corresponding to two load impedances. The switching of RF transistor between the two states for the BPSK modulation is performed with the help of a micro-controller. A typical BSN (see Fig. \ref{fig:tag}) circuit consists of an antenna, transmitter, receiver, energy harvester, variable impedances, micro-controller, sensor, RF transistor switch, and a battery.

\subsection{Channel~Model}
We consider an uplink transmission scenario in which each cluster of two BSNs is served by a single CE and BSR\@. Both BSNs and BSRs are equipped with single antennas. As shown in Fig. \ref{fig:System_model}, \(h_{f,i}\) is used to denote the channel coefficient of the forward link between the CE and \(i\)th BSN, for \(\,i  \in\{1,2\}\), whereas \(h_{b,i}\) is used to denote the  channel coefficient of the backscatter link between \(i\)th BSN and BSR. 

For the \emph{forward channel}, path loss only propagation effects are considered in \(h_{f,i}\). This fading-free channel model is a reasonable channel assumption because of the proximity and strong line-of-sight (LoS) link between CE and BSNs\footnote{A fading impaired channel model for forward link is left as a future study at the moment.}. Combined with our BackCom model and given that the transmit power of the CE is \(P_{T}\), the power required at a BSN for forward channel is given by \(P_{T}\,\varGamma_{i}\,|h_{f,i}|^2\). 
In the \emph{backscatter channel}, we assume that the channel coefficient \(h_{b,i}\) follow independent and non-identically distributed (i.n.i.d) Nakagami-\(m\) fading which can model both LoS and non-LoS conditions. The zero mean additive white Gaussian noise (AWGN), \(w\), with variance \(N_0\), i.e., \(w\sim \mathcal{C}\mathcal{N}\)(0, \(N_0\)) is also considered in the system.

\subsection{NOMA Scheme}
We employ a power-domain NOMA (PD-NOMA) uplink scheme in the BackCom system. PD-NOMA operates by multiplexing users with relatively large channel gain differences over same time/frequency slot, thereby improving spectral efficiency. During IoT deployment, sometimes grouping IoTs with nearly balanced power differences is inevitable. However, in IoT scenarios where conventional PD-NOMA is unusable, its BackCom variant can still function by manipulating the power reflection coefficients of the BSNs to generate acceptable channel differences. In a cluster of two BSNs, a training phase is required by the BackCom to differentiate the BSNs into a weak or strong user. It works as follows: each BSN is distinguishable by its unique ID which is known to the BSR. The BSR broadcasts a pilot signal with the unique ID for each BSN in two training time slots. The BSNs backscatter the signal with the same power reflection coefficient after receiving its unique ID. Now, the BSR can obtain the instantaneous channel state information (CSI) and can classify the BSNs into weak/strong pair. The BSNs can then set its power reflection coefficients corresponding to the received instantaneous CSI by the BSR.

For a single cluster, the received signal at the BSR can be written as
\begin{equation}\label{eq:composite}
  y = \sqrt{P_T \varGamma_1} h_{f,1} h_{b,1} x_1 + \sqrt{P_T \varGamma_2} h_{f,2} h_{b,2}  x_2 + w,
\end{equation}
where \(x_i\) is the BPSK modulated information signal of $i$th nodes. As mentioned earlier, the forward channels \(h_{f,i}\)s are assumed to be dominated by a deterministic path loss model only, whereby their effects are compensated in the transmit SNR of both BSNs and are thus omitted in further analysis. In order to decode the signals transmitted by the BSNs, successive interference cancellation (SIC) process is implemented at the BSR, where error propagation may happen. Without the loss of generality, it assumed that the first BSN has higher channel gain than the second BSN, i.e., \(|h_{b,1}|^2 > |h_{b,2}|^2 \). To detect the weaker signal in a SIC scheme, the stronger  signal should be detected first and scaled, then subtracted from the aggregate received signal. Therefore, the optimal decoding order is in the order of decreasing channel gains. Considering this, the signal from BSN-1, $u_1$, is decoded firstly by treating BSN-2, $u_2$, as inter-user interference (IUI) at the BSR. Thus the maximum likelihood detector (MLD) of $u_1$, given that the channel gains are estimated perfectly at BSR, can be described by \cite{proakis2007digital}
\begin{equation}\label{eq:MLD}
  \widehat{x}_1  = 
  \underset{\widetilde{x}_1 \in \mathcal{S}}{\operatorname{arg\,min}} \bigg| y- \sqrt{P_T \varGamma_1} h_{b,1} \widetilde{x}_1\bigg|^2,
\end{equation}
where $\widehat{x}_1 $ is the estimated data symbol, $\mathcal{S}$ is the set of all possible constellation points for \(u_1\), and $\widetilde{x}_1$ is the set of all possible trial values for \(x_1\).

Next, $u_2$ signal is decoded after subtracting the detected $u_1$ symbol from received composite signal \(y\). If $u_1$'s signal is decoded correctly then no IUI is faced in decoding of $u_2$, otherwise, there will be error propagation from $u_1$ while decoding $u_2$. The detector for $u_2$ can be expressed as
\begin{equation}\label{eq:MLD1}
  \widehat{x}_2  = 
  \underset{\widetilde{x}_2 \in \mathcal{S}}{\operatorname{arg\,min}} \bigg| \bigg(y- \sqrt{P_T \varGamma_1} h_{b,1} \widehat{x}_1 \bigg) - \sqrt{P_T \varGamma_2} h_{b,2} \widetilde{x}_2 \bigg|^2.
\end{equation}
 
In the next section, BER expressions are derived for PD-NOMA-BackCom system with a cluster of two BSNs, assuming equiprobable symbols. The same approach may be applied for a higher order phase shift keying (PSK) modulation, however, the derivation becomes impractical when a large number of BSNs are multiplexed together.

\section{NOMA-BackCom BER Performance Analysis}

\begin{figure}[ht]
  \centering
  \includegraphics[width=\columnwidth]{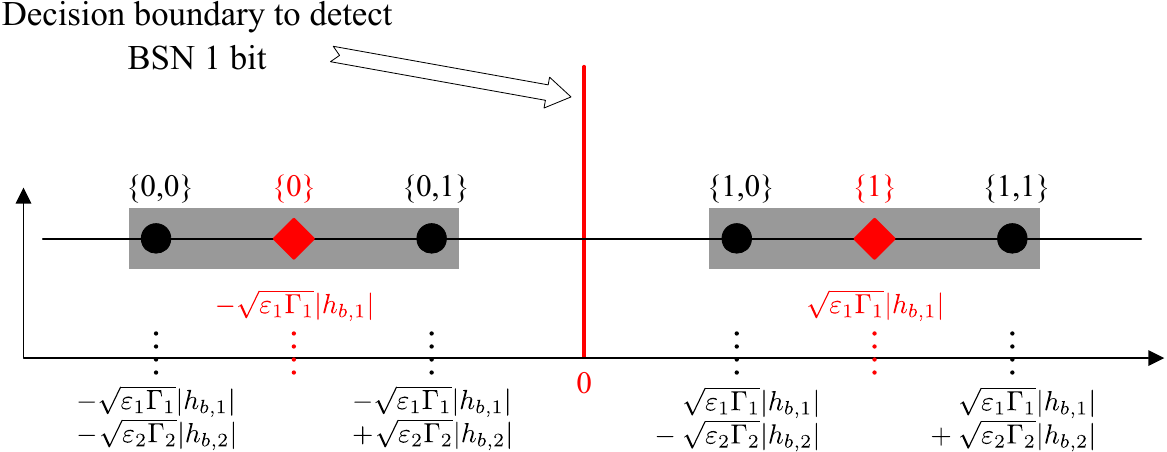}
  \caption{The received signal space diagram  of super-imposed BPSK symbols from two BSNs at the BSR, where \(|h_{b,1}|^2 > |h_{b,2}|^2 \).}
  \label{fig:constellation}
\end{figure}

As noted from (\ref{eq:composite}), the received symbol at BSR is a superposition of two BPSK symbols, therefore, it corresponds to a total of four constellation points as shown in Fig. \ref{fig:constellation}. Because \(|h_{b,1}|^2 > |h_{b,2}|^2 \), the BSR detects \(u_1\) symbol first. Thereafter, it subtracts the decoded \(u_1\) symbol from the received signal and detects \(u_2\) symbol. Each constellation point in Fig. \ref{fig:constellation} is represented by two bits given by \(\{x_1, x_2\}\). Here, \(x_1\) is the BPSK bit of $u_1$ and \(x_2\) is the BPSK bit of $u_2$. In Fig. \ref{fig:constellation}, red diamond shows the original location of \(x_1\) without IUI from $u_2$. However, because of the presence of IUI from $u_2$, \(x_1\) is translated to two possible constellation points. The shaded block shows the two possible values that a particular $u_1$ bit \(x_1\) may take due to IUI. By modifying the values of reflection coefficients, the position of these constellation points can be changed, which will in turn affect the BER performance of the system.

\subsection{BER of the First User}
The detection of the first user is performed according to (\ref{eq:MLD}), therefore, no SIC is required in the process. An error will be made in the detection process if $\widehat{x}_1\neq  x_1 $ and its probability is given by $P_{e}(u_1)$. The probability of error for $u_1$ depends on the decision boundary distance of $x_1$ and it is the sum of error probabilities of each possible symbol multiplied with the prior probability. Because of IUI, there are four possible cases in which a bit $x_1$ can be decoded incorrectly. When \(u_1^{(0)}\) and \(u_2^{(0)}\) are sent, where \(u_i^{(y)}\) represents that user $i$ transmits a bit $y$, the decision boundary of $x_1$, represented by the red line, is at a distance of \( \sqrt{\varepsilon_1 \Gamma_1}|h_{b,1}| + \sqrt{\varepsilon_2 \Gamma_2 }|h_{b,2}| \) from the constellation point \{0,\! 0\}, where $\varepsilon_i$ denote the symbol energy of user $i$. Therefore, bit $x_1$ will be decoded incorrectly if in-phase component of $w$ exceeds the sum of $u_1^{(0)}$ and $u_2^{(0)}$ i.e., $w \geq  \sqrt{\varepsilon_1 \Gamma_1}|h_{b,1}| + \sqrt{\varepsilon_2 \Gamma_2 }|h_{b,2}| $. Similarly, following the same procedure for each symbol, the error is calculated and multiplied by prior probability. Consequently, the error probability for $u_1$, considering even symmetry in constellation diagram, is given by
\begin{equation}\label{eq:P1e}
  \begin{split}
    P_{e}(u_1)  =  \frac{1}{4} \left[  \mathbb{P}(|w|\geq \Upsilon_1  + \Upsilon_2  )  \right. 
    + \left. \mathbb{P}(|w|\geq \Upsilon_1 - \Upsilon_2  )  \right],
  \end{split}
\end{equation}
where $\mathbb{P}(x)$ denotes the probability of event $x$, and $\Upsilon_i=\sqrt{\varepsilon_i \Gamma_i}|h_{b,i}|$.
The expression (\ref{eq:P1e}) can be represented using Gaussian $\mathcal{Q} $ function as
\begin{equation}\label{eq:P1e_Q}
  \begin{split}
    P_{e}(u_1)  =  \frac{1}{2} \left[ \mathcal{Q}\left(Y \right) + \mathcal{Q}\left(Z \right)  \right],
  \end{split}
\end{equation}
where $Y$ and $Z$ are random variables (RVs), defined as
\begin{equation}\label{eq:Y_Z}
  \begin{split}
    Y = \frac{\Upsilon_1}{\sqrt{N_0/2}} + \frac{\Upsilon_2}{\sqrt{N_0/2}}, \\
    Z = \frac{\Upsilon_1}{\sqrt{N_0/2}} - \frac{\Upsilon_2}{\sqrt{N_0/2}},
  \end{split}
\end{equation}
and the Gaussian $\mathcal{Q} $ function is defined as
\begin{equation}\label{eq:Qfunc}
  \mathcal{Q}(x) = \frac{1}{2\pi} \int_{x}^{\infty}\exp\bigg(\!-\frac{u^2}{2}\bigg)  \,\mathrm{d}u.
\end{equation}

The average BER for BSN-1, denoted as $\overline{P_{e}(u_1)}$, is evaluated by averaging over the PDF of RV $Y$, $f_{Y}(y)$, and RV $Z$, $f_{Z}(z)$, where, $f_{Y}(y)$ and $f_{Z}(z)$ are the PDFs for the sum and difference of two i.n.i.d Nakagami-$m$ distributions, respectively. The average error probability of $u_1$ is 
\begin{equation}\label{eq:P1e_Q1}
  \begin{split}
    \overline{P_{e}(u_1) }  =  \frac{1}{2} \left[ \int_{-\infty}^{\infty }   \mathcal{Q}\left(Y\right)\! f_{Y}(y) \,\mathrm{d}y + \right. 
    \left. \int_{-\infty}^{\infty }  \mathcal{Q}\left(Z \right)\! f_{Z}(z) \,\mathrm{d}z \right].
  \end{split}
\end{equation}

The PDF $f_{Y}(y)$ has been derived in \cite{Annamalai2000} and \cite{Rahman2010} in the form of Appell hypergeometric function of the second kind and Lauricella multivariable hypergeometric function, respectively. However, a closed-form solution of (\ref{eq:P1e_Q1}) is prohibited using expressions in \cite{Annamalai2000} and \cite{Rahman2010}, rendering the use of numerical evaluation. Therefore, the PDF of $Y$ in this work is approximated with another distribution, complying with the features of $Y$ as shown in the following lemma.

\emph{Lemma 1:} The distribution of the sum of two i.n.i.d Nakagami-$m_i$ RVs, $Y_1$ and $Y_2$, with parameters $m_i$ and $\Omega_i$, where \(\,i  \in\{1,2\}\),  can be approximated by another Nakagami-$m_{R_1}$ RV $R_1$, i.e., $R_1=Y_1+Y_2$, with fading parameter $m_{R_1}$ and average power $\Omega_{R_1}$, with PDF as
\begin{equation}\label{eq:PDF_nak}
  f_{R_1}(r) =  \frac{2m_{R_1}^{m_{R_1}}r^{2m_{R_1}-1}}{\Omega_{R_1}^{m_{R_1}}\Gamma (m_{R_1})}\text{exp}(-\frac{m_{R_1}r^2}{\Omega_{R_1}} ) ,\quad r_1> 0,
\end{equation}
where $\Omega_{R_1}$ and $m_{R_1}$ are defined as \cite{NAKAGAMI19603}
\begin{equation}\label{eq:m_O}
  \begin{split}
    \Omega_{R_1} &= \mathbb{E}[{R_1}^2],\\
    m_{R_1} = \frac{(\mathbb{E}[{R_1}^2])^2}{Var[{R_1}^2]} &= \frac{\Omega_{R_1}^2 }{\mathbb{E}[{R_1}^4]-\Omega_{R_1}^2},
  \end{split}  
\end{equation}
where $\Gamma(.)$ is the gamma function, and $\mathbb{E}[.]$ and $Var[.]$ denote the expectation and variance operators, respectively.

\emph{Proof:} A two-step process is followed to approximate the distribution of the sum of two RVs $Y_1$ and $Y_2$ to another distribution. In the first step, the method of moments (MoM) approach is utilized to match the moments of RV $Y=Y_1+Y_2$ to that of Nakagami-$m_{R_1}$ RV, ${R_1}$. Nakagami-$m$ distribution is based on two parameters, i.e., fading parameter $m$ and shape parameter $\Omega$. Therefore, we need to match the moments of the two parameters. From \cite{NAKAGAMI19603}, the $n$th moment of the Nakagami-$m_i$ distribution is given by
\begin{equation}\label{eq:nmoments}
  \mathbb{E}[Y_i^n] = \frac{\Gamma(m_i+\frac{n}{2})}{\Gamma(m_i)} \bigg(\frac{\Omega_i}{m_i}\bigg)^{\frac{n}{2}},
\end{equation}
where $m_i$ is the fading parameter of each RV $Y_i$, $\Omega_{1}=\Omega_{Y_1}=2 \Upsilon_1^2/N_0$ and $\Omega_{2}=\Omega_{Y_2}=2 \Upsilon_2^2/N_0$. Using the multinomial theorem assuming independence among RVs, $Y_1$, $Y_2$, and (\ref{eq:nmoments}), the second moment of the sum RV $Y=Y_1+Y_2$ is matched with the second moment of RV $R_1$, i.e.,
\begin{align}\label{eq:2moment}
    &\Omega_{R_1}={\mathbb{E}[Y^2]} = \mathbb{E}[Y_1^2] +\mathbb{E}[Y_2^2]+2\mathbb{E}[Y_1]\mathbb{E}[Y_2]\nonumber\\
    &= \Omega_1 + \Omega_2+ 2 \frac{\Gamma(m_1+\frac{1}{2})}{\Gamma(m_1)} \bigg(\frac{\Omega_1}{m_1}\bigg)^{\frac{1}{2}} \frac{\Gamma(m_2+\frac{1}{2})}{\Gamma(m_2)} \bigg(\frac{\Omega_2}{m_2}\bigg)^{\frac{1}{2}} .
\end{align}

Now we need to find the value of $m_{R_1}$. To this objective, we note from (\ref{eq:m_O}) that $\mathbb{E}[{R_1}^4]$ is required. Again using the multinomial theorem for ${R_1}=Y_1+Y_2$, $\mathbb{E}[{R_1}^4]$ can be evaluated as 
%
\begin{equation}\label{eq:4moment}
    \mathbb{E}[{R_1}^4] = \mathbb{E}[Y_1^4] + 4\mathbb{E}[Y_1^3]\mathbb{E}[Y_2] +6\mathbb{E}[Y_1^2]\mathbb{E}[Y_2^2]
     +4\mathbb{E}[Y_1]\mathbb{E}[Y_2^3] +\mathbb{E}[Y_2^4].
\end{equation}
Then, from (\ref{eq:m_O}) and (\ref{eq:2moment}), value of $m_{R_1}$ can easily be found.

In the second step, the Kolmogorov-Smirnov (K-S) test is used for goodness-of-fit (see Appendix A). 
Fig. \ref{fig:ECDF}\ref{sub@fig:ECDF_Nakagami} shows the plot of empirical CDF and approximated theoretical CDF.

\begin{figure}[!t]
  \centering
  \vspace{-10pt}
  \subfloat[]{\includegraphics[width= 3.35in]{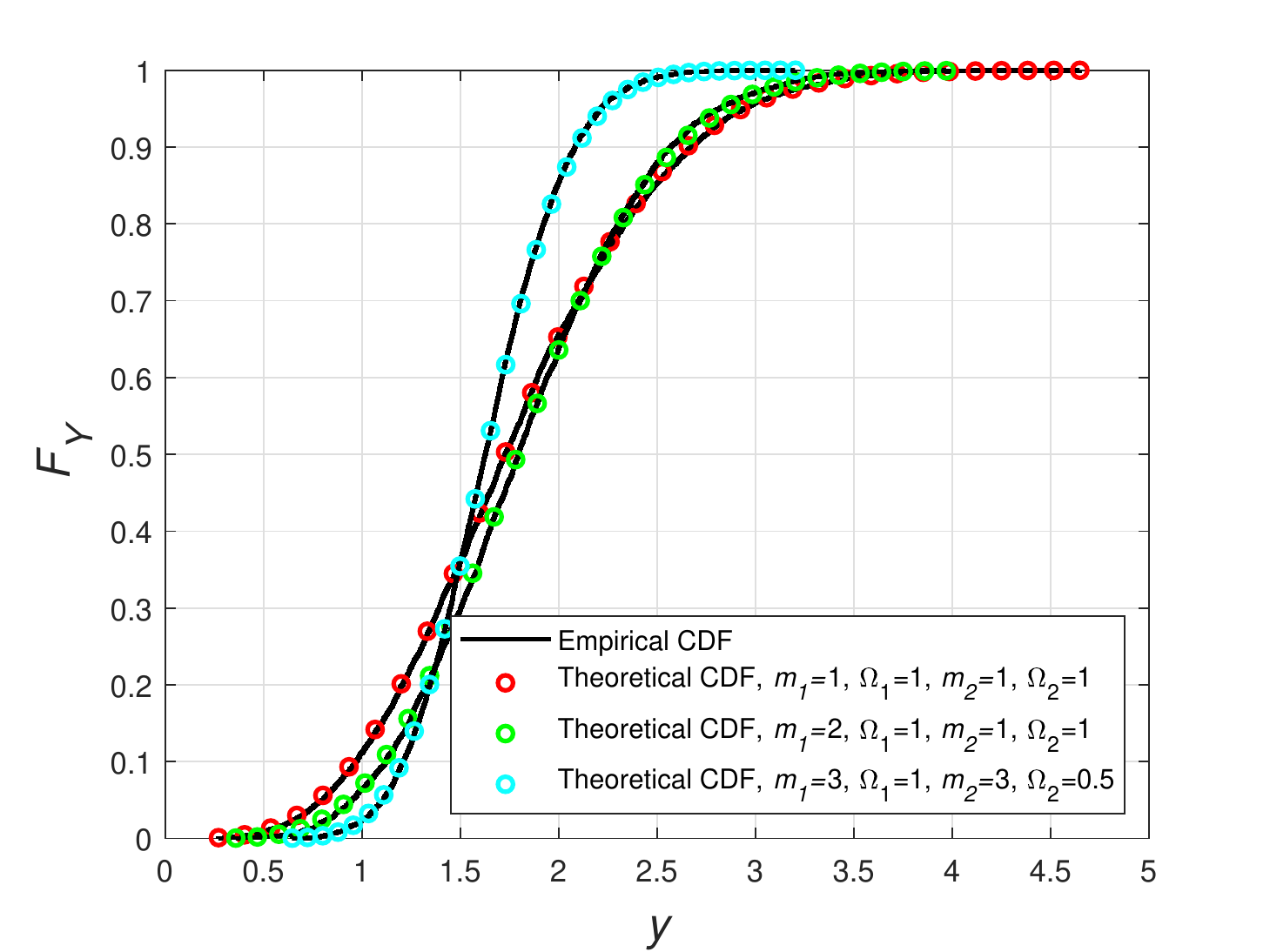}\label{fig:ECDF_Nakagami}}
  \subfloat[]{\includegraphics[width= 3.35in]{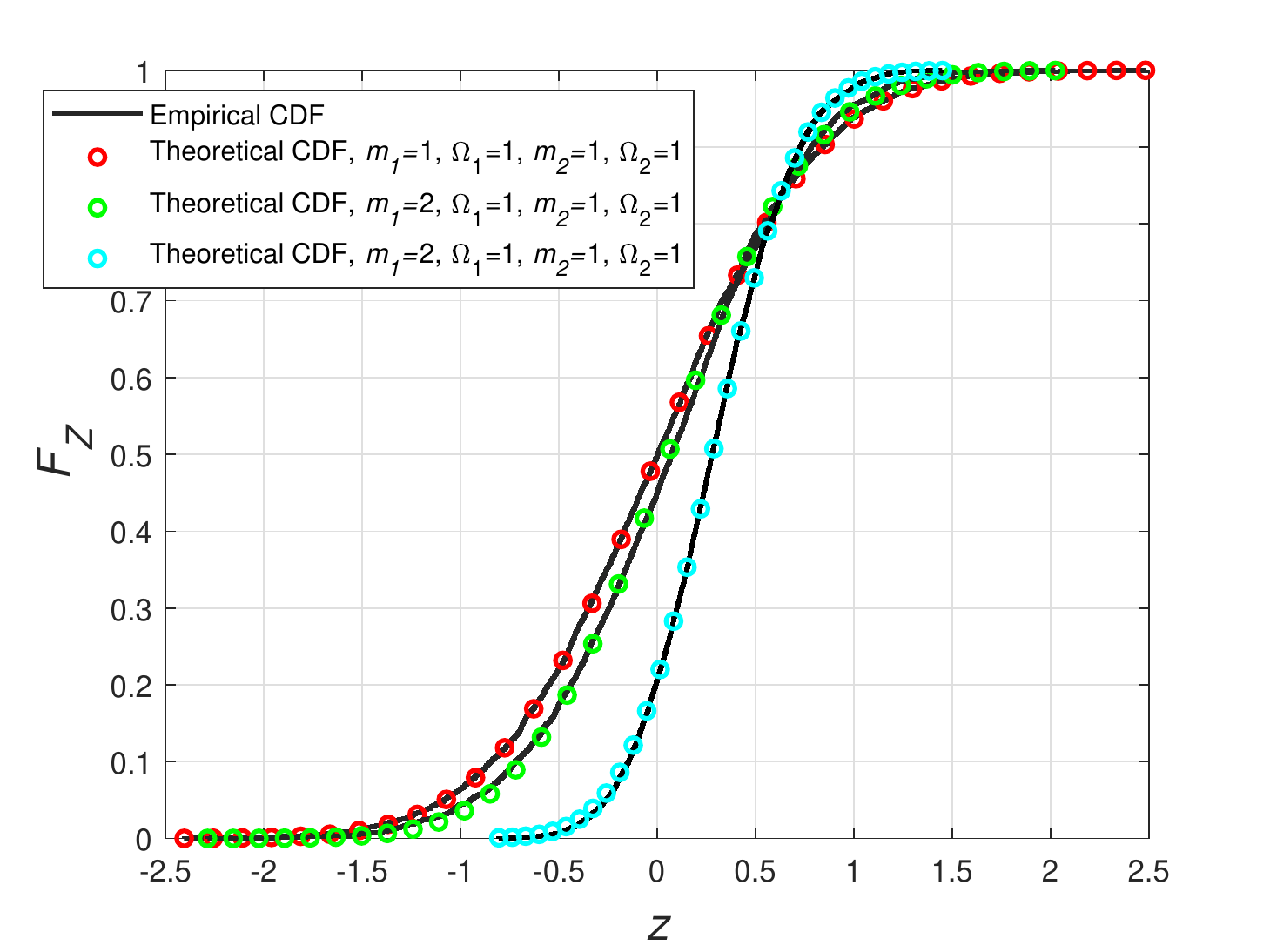}\label{fig:ECDF_Normal}}
  \caption{Empirical CDF vs theoretical CDF obtained using MOM with two i.n.i.d Nakagami-$m$ distributions; (a) the sum of the two, (b) the difference of two.}
  \label{fig:ECDF}
\end{figure}


The PDF $f_{Z}(z)$ for the difference of two i.n.i.d Nakagami-$m$ distributions has not been derived previously. Therefore, we approximate it with another closely matching distribution using the following lemma.

\emph{Lemma 2:} The distribution of the difference of two i.n.i.d Nakagami-$m_i$ RVs, $Z_1$ and $Z_2$, with parameters $m_i$ and $\Omega_i$, where \(\,i  \in\{1,2\}\), can be approximated by a Gaussian (normal) RV, ${W_1}=Z_1-Z_2$, with mean $\mu_{W_1}$ and variance $\sigma^2_{W_1}$, with PDF 
\begin{equation}\label{eq:normal_PDF}
  f_{W_1}(w) = \frac{1}{\sqrt{2\pi\sigma_{W_1}^2}}\exp\bigg(-\frac{(w-\mu_{W_1})^2}{2\sigma^2_{W_1}}\bigg),
\end{equation}
where the values of $\mu_{W_1}$ and $\sigma^2_{W_1}$ are given in (\ref{normal_1_mom}) and (\ref{normal_2_mom}).

\emph{Proof:}  A similar two-step process as used in Lemma 1, is used to approximate the distribution of the difference of two RVs $Z_1$ and $Z_2$ to another distribution. Firstly, MoM is applied to match the moments of RV $Z=Z_1-Z_2$ to moments of the normal RV, ${W_1}$. The first two moments of ${W_1}$ are
 \begin{equation}\label{normal_moments}
   \begin{split}
     \mathbb{E}[{W_1}] &= \mu_{W_1},\\
     \mathbb{E}[{W_1}^2] &= \mu_{W_1}^2+\sigma_{W_1}^2.
   \end{split}
 \end{equation}

Following the algorithm of moment matching, using multinomial theorem, noting independence among RVs, $Z_1$ and $Z_2$, and using (\ref{eq:nmoments}), the mean $\mu_{W_1}$ of the normal RV, ${W_1}$, can be evaluated by matching the first moment of the difference RV $Z=Z_1-Z_2$  with RV ${W_1}$ as
\begin{align}\label{normal_1_mom}
    \mu_{W_1} &= \mathbb{E}[Z] = \mathbb{E}[Z_1] -\mathbb{E}[Z_2]\nonumber\\
    &= \frac{\Gamma(m_1+\frac{1}{2})}{\Gamma(m_1)} \bigg(\frac{\Omega_1}{m_1}\bigg)^{\frac{1}{2}} - \frac{\Gamma(m_2+\frac{1}{2})}{\Gamma(m_2)} \bigg(\frac{\Omega_2}{m_2}\bigg)^{\frac{1}{2}},
\end{align}
where $\Omega_{1}=\Omega_{Z_1}=2 \Upsilon_1^2/N_0$ and $\Omega_{2}=\Omega_{Z_2}=2 \Upsilon_2^2/N_0$. Now we match the second moments to obtain variance $\sigma_{W_1}^2$
\begin{align}\label{normal_2_mom}
    \mu_{W_1}^2+\sigma_{W_1}^2 &=\mathbb{E}[Z^2] = \mathbb{E}[Z_1^2] +\mathbb{E}[Z_2^2] -2\mathbb{E}[Z_1]\mathbb{E}[Z_2],\nonumber\\
    \sigma_{W_1}^2 &= \mathbb{E}[Z^2] - \mu_{W_1}^2,
\end{align}
where $\mathbb{E}[Z_i^n]$ is taken from (\ref{eq:nmoments}). 

In the second step, the Kolmogorov-Smirnov (K-S) test is used to show that the normal distribution closely matches the distribution of the difference of two Nakagami-$m$ RVs. The K-S test is 
has been performed in Appendix B.
 Fig. \ref{fig:ECDF}\ref{sub@fig:ECDF_Normal} shows the plot of empirical CDF and approximated theoretical CDF showing a close agreement.


Now from (\ref{eq:PDF_nak}) and (\ref{eq:normal_PDF}), a closed-form solution for the average BER can be evaluated as follows
%
\begin{equation}\label{eq:Pu1_avergae}
  \overline{P_{e}(u_1) } = \frac{1}{2} \Big[ \underbrace{\int_{-\infty}^{\infty }\!\!\!\!\! \mathcal{Q}(R) f_{R }(r ) \,\mathrm{d}r}_{\Phi}  + \underbrace{\int_{-\infty }^{\infty } \!\!\!\!\!  \mathcal{Q}\left(W \right)\! f_{W}(w) \,\mathrm{d}w}_{\Lambda }  \Big].
\end{equation}
The closed-form expression for $\Phi$ can be found by invoking the alternative form of the Q-function known as Craig's formula \cite{Simon2005} and applying moment generating function (MGF) for $f_{\gamma_R }(\gamma_r )$. It is given by 
\begin{equation}\label{eq:Nak_MGF}
  \begin{split}
    \Phi(m,\overline{\gamma}) = 
    \begin{cases}
      \frac{1}{2} \Bigg[ 1 - \Psi \big(\frac{\overline{\gamma}}{2m}\big) \sum\limits_{k = 0}^{m-1}\bigg(\frac{1-\Psi^2\big(\frac{\overline{\gamma}}{2m}\big)^k}{4}\bigg) \Bigg],
       \text{$m$ integer}\\
      \frac{1}{2\sqrt{\pi}}\frac{\sqrt{\frac{\overline{\gamma}}{2m}}}{(1+\frac{\overline{\gamma}}{2m})^{m+(1/2)}} \frac{\Gamma(m+\frac{1}{2})}{\Gamma(m+1)} \times \\
      {}_{2}F_{1} \bigg( 1,m+\frac{1}{2};m+1;\frac{m}{m+\frac{\overline{\gamma}}{2m}}\bigg),
      \: \text{$m$ noninteger}
    \end{cases}       
  \end{split}
\end{equation}
where $\Psi\big(\frac{\overline{\gamma}}{2m}\big)\triangleq \sqrt{\frac{\overline{\gamma}/2}{m+\overline{\gamma}/2}}$ and ${}_{2}F_{1}$ 
is the Gauss hypergeometric function. Further, the closed-form expression for $\Lambda$ in (\ref{eq:Pu1_avergae}) is given by following Lemma.

\emph{Lemma 3:} The integral of the product of a Q function and a normal distribution with mean $\mu$ and variance $\sigma^2$ is 
\begin{equation}\label{eq:normal_closed}
 \Lambda(\mu,\sigma^2) = \mathcal{Q}\bigg(\frac{\mu}{\sqrt{\sigma^2+1}}\bigg).
\end{equation}

\emph{Proof:} See Appendix C.

Finally by using (\ref{eq:Nak_MGF}) and (\ref{eq:normal_closed}) in (\ref{eq:Pu1_avergae}), the average probability of error for BSN-1 can be expressed as
\begin{equation}\label{eq:Pu1_average_closed}
  \overline{P_{(e)}(u_1) } = \frac{1}{2} \left[ \Phi(m_{R_1},\Omega_{R_1}) + \Lambda(\mu_{W_1},\sigma^2_{W_1})\right] .
 \end{equation}

\subsection{BER of the Second User}

To decode the second user, SIC process is implemented, whereby, BSR would initially detect the first user according to (\ref{eq:MLD}) and then from the reconstructed $\widehat{x}_1$ bit, compute, 
\begin{equation}\label{MLD_2ndSIC}
  \begin{split}
    \widehat{x}_2= \underset{\widetilde{x}_2 \in \mathcal{S}}{\operatorname{arg\,min}} \bigg| y_{SIC} + w - \sqrt{P_T \varGamma_2} h_{b,2} \widetilde{x}_2\bigg|^2.
  \end{split}
\end{equation}

\begin{figure}[ht]
  \centering
  \includegraphics[width=0.8\columnwidth]{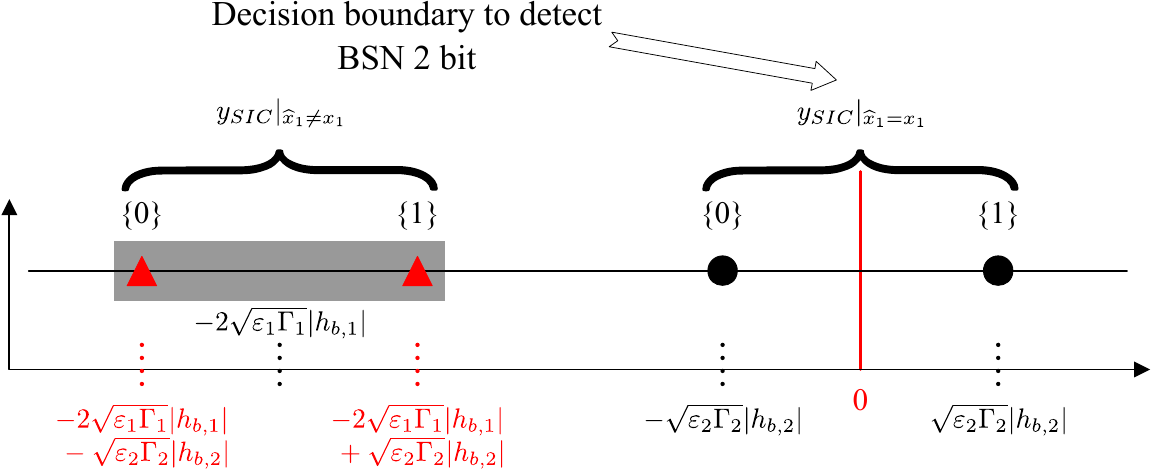}
  \caption{The received signal space diagram of BSN-2 for $x_1^{(1)}$ transmission when $y_{SIC}|_{\widehat{x}_1\neq  x_1}$ and $y_{SIC}|_{\widehat{x}_1=  x_1}$.}
  \label{fig:Constellation-I_II}
\end{figure}

Therefore, if $\widehat{x}_1$ is detected correctly, i.e., $\widehat{x}_1= x_1$, then $y_{SIC} = \sqrt{P_T \varGamma_2} h_{b,2} x_2$ and it corresponds to an IUI-free decoding. On the other hand, if $\widehat{x}_1$ is detected incorrectly, i.e., $\widehat{x}_1\neq  x_1$, then $y_{SIC} = \sqrt{P_T \varGamma_1} h_{b,1}x_1 + \sqrt{P_T \varGamma_2}h_{b,2} x_2 - \sqrt{P_T \varGamma_1}h_{b,1} \widehat{x}_1 $ and the decoding decision for second user also depends on the value of reconstructed bit $ \widehat{x}_1$. Fig. \ref{fig:Constellation-I_II} shows the received signal diagram for the two aforementioned cases when the transmitted $u_1$ bit $x_1^{(1)}$ is decoded correctly or incorrectly. Therefore, these two cases should be handled differently.

\subsubsection{Case I}

We first consider the case that $u_1$ symbol has been detected correctly by BSR. The probability for correct decoding of $x_1$ bit is opposite to the one derived in (\ref{eq:P1e}), which gives the probability for incorrect decoding of $x_1$ bit. This will serve as the prior probability for case I. The error probability for $x_2$ bit is influenced by the decision boundary for detecting $u_2$ symbol as shown by the constellation points indicated by the black circles in Fig. \ref{fig:Constellation-I_II}. Hence, the probability of error for $u_2$ with correct $u_1$ decoding, considering both $x_1^{(0)}$ and $x_1^{(1)}$ bit transmission scenarios, is given by (\ref{eq:P_e_u2_I}) at the top of page.

The conditional expressions pertain to decoding the $x_1$ bit correctly and can be inferred from Fig. \ref{fig:constellation}. The first and second terms in (\ref{eq:P_e_u2_I}) correspond to $x_2^{(0)}$ and $x_2^{(1)}$ received constellation points, respectively.
%

\begin{figure*}[!t]
  \vspace{-10pt}
  \normalsize
  \setcounter{MYtempeqncnt}{\value{equation}}
  \setcounter{equation}{22}
  \begin{multline}
  \label{eq:P_e_u2_I}
  P_e^I(u_2) = \frac{1}{2} \left[ \mathbb{P}(w\leq \Upsilon_1 + \Upsilon_2 ) \times \mathbb{P}\big(w\geq   \Upsilon_2\bigm\vert w\leq \Upsilon_1 + \Upsilon_2 \big) +  \mathbb{P}(w\leq \Upsilon_1 - \Upsilon_2 ) \times \right. \\ \left.\mathbb{P}\big(w\leq -\Upsilon_2\bigm\vert w\leq \Upsilon_1 - \Upsilon_2 \big)\right].
  \end{multline}
  \vspace{-10pt}
  \begin{multline}\label{eq:P_e_u2_II}
  P_e^{II}(u_2) =  \frac{1}{2} \left[ \mathbb{P}(w\geq \Upsilon_1 + \Upsilon_2 ) \times  \mathbb{P}\big( w\geq 2\Upsilon_1 + \Upsilon_2 \bigm\vert w\geq \Upsilon_1 + \Upsilon_2 \big)+ \mathbb{P}(w\geq \Upsilon_1 - \Upsilon_2 )\times \right. \\ \left. \mathbb{P}\big( w\leq 2\Upsilon_1 - \Upsilon_2 \bigm\vert w\geq \Upsilon_1 - \Upsilon_2 \big) \right].
  \end{multline}
  \setcounter{equation}{\value{equation}}
  \hrulefill
  \vspace*{-4pt}
\end{figure*}

Using conditional probability law, (\ref{eq:P_e_u2_I}) can be rewritten as
\begin{equation}\label{eq:P_e_u2_I_simplified}
  P_e^I(u_2) = \frac{1}{2}\left[ \mathbb{P}(\Upsilon_2 \leq w \leq \Upsilon_1 + \Upsilon_2) + \mathbb{P}(\Upsilon_2 \leq w)\right].
\end{equation}

The  expression in (\ref{eq:P_e_u2_I_simplified}) can be  represented  using  Gaussian $\mathcal{Q}$ function as
\begin{equation}\label{eq:P_e_u2_I_Q}
  P_e^I(u_2) = \mathcal{Q}(\sqrt{\gamma_{\Upsilon_2}}) - \frac{1}{2}\mathcal{Q}(Y),
\end{equation}
where $Y$ is defined in (\ref{eq:Y_Z}) and $\gamma_{\Upsilon_2} = \frac{\Upsilon_2^2}{N_0/2}$.

\subsubsection{Case II}
This case considers the scenario when $u_1$ symbol has been detected incorrectly by BSR. The probability for incorrect decoding of $x_1$ bit is similar to the one derived in (\ref{eq:P1e}). This will serve as the prior probability for case II. Because of the IUI from $u_1$, there will be an error propagation of $2\Upsilon_1$ from the decision boundary to detect $x_2$. This can be observed from the constellation points represented by the red triangles in Fig. \ref{fig:Constellation-I_II}. Hence, the probability of error for $u_2$ with incorrect $u_1$ decoding, considering both $x_1^{(0)}$ and $x_1^{(1)}$ bit transmission scenarios, is (\ref{eq:P_e_u2_II}).

The conditional expressions in (\ref{eq:P_e_u2_II}) are obtained in the same way as in (\ref{eq:P_e_u2_I}) except that $x_1$ is decoded incorrectly in case II. By using conditional probability law, the probability of error for $u_2$ considering $x_1$ is decoded erroneously, represented using the Gaussian $\mathcal{Q}$ function, is given by
\begin{equation}\label{eq:P_e_u2_II_Q}
  P_e^{II}(u_2) = \frac{1}{2} \left[ \mathcal{Q}(C) +\mathcal{Q}(Z) -\mathcal{Q}(D)\right].
\end{equation}
where $Z$ is defined in (\ref{eq:Y_Z}), $C$ and $D$ are RVs, defined as 
\begin{equation}\label{eq:C_D}
  \begin{split}
    C = \frac{2\Upsilon_1}{\sqrt{N_0/2}} + \frac{\Upsilon_2}{\sqrt{N_0/2}}, \\
    D = \frac{2\Upsilon_1}{\sqrt{N_0/2}} - \frac{\Upsilon_2}{\sqrt{N_0/2}}.
   \end{split}
 \end{equation}
Then, the total probability of error for $u_2$ can be found as the sum of both the cases given by (\ref{eq:P_e_u2_I_Q}) and (\ref{eq:P_e_u2_II_Q})
\begin{equation}\label{eq:P_e_u2_Q}
  \begin{split}
    P_e(u_2) &= P_e^I(u_2) + P_e^{II}(u_2), \\
    P_e(u_2) = \mathcal{Q}(\sqrt{\gamma_{\Upsilon_2}}) &+ \frac{1}{2}\left[-\mathcal{Q}(Y) + \mathcal{Q}(C) +\mathcal{Q}(Z) -\mathcal{Q}(D)\right].\nonumber
  \end{split}
\end{equation}

The RV $C$, i.e., sum of two i.n.i.d Nakagami-$m_i$ RVs, can be approximated by a Nakagami-$m_{R_2}$ RV with fading parameter $m_{R_2}$ and average power $\Omega_{R_2}$, with PDF given in (\ref{eq:PDF_nak}) as proved in Lemma 2, where $\Omega_{C_1}=4 \Omega_{1}$ and $\Omega_{C_2}=\Omega_{2}$ . The RV $D$, i.e., difference of two i.n.i.d Nakagami-$m_i$ RVs, can be approximated by a Gaussian (normal) RV, with mean $\mu_{W_2}$ and variance $\sigma^2_{W_2}$, with PDF given in (\ref{eq:normal_PDF}) as proved in Lemma 2, where $\Omega_{D_1}=4 \Omega_{1}$ and $\Omega_{D_2}=\Omega_{2}$ .

Therefore, the average probability of error for $u_2$ becomes
\begin{multline}\label{eq:Pu2_average}
  \overline{P_e(u_2) } = \Phi(m_2, \gamma_{\Upsilon_2}) +\frac{1}{2}\left[ -\Phi(m_{R_1},\Omega_{R_1}) +\Phi(m_{R_2},\Omega_{R_2}) 
  +  \Lambda(\mu_{W_1},\sigma^2_{W_1})-\Lambda(\mu_{W_2},\sigma^2_{W_2}) \right].
\end{multline}


\section{Numerical Results and Simulations}

\begin{figure*}[h]
  \centering
  \subfloat[$\Gamma_1=1, \Gamma_2=0.9$]{\includegraphics[width= 3.35in]{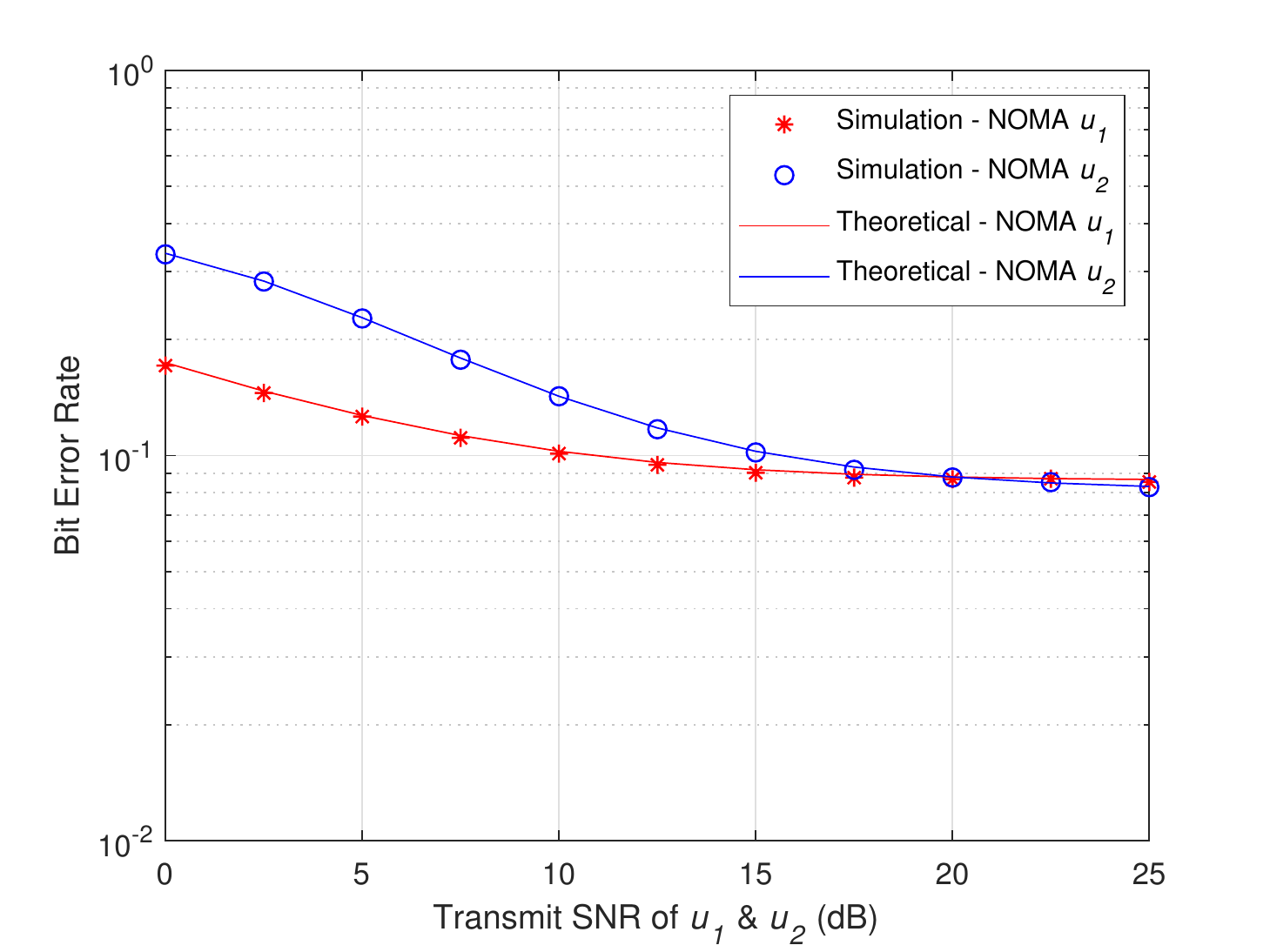}\label{fig:BER_plots_1}}
  \subfloat[$\Gamma_1=1, \Gamma_2=0.6$]{\includegraphics[width= 3.35in]{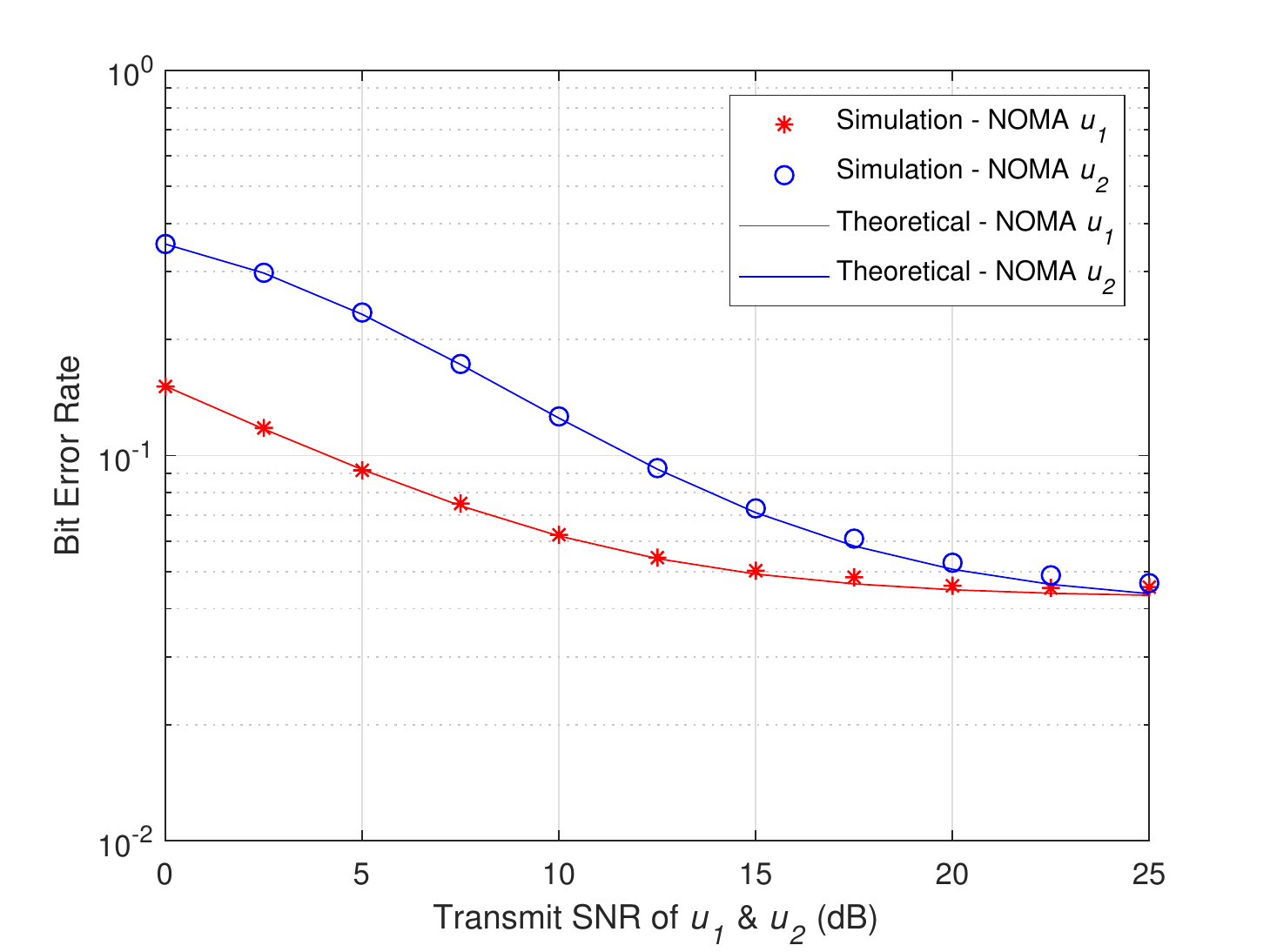}\label{fig:BER_plots_2}}\\
  \subfloat[$\Gamma_1=1, \Gamma_2=0.3$]{\includegraphics[width= 3.35in]{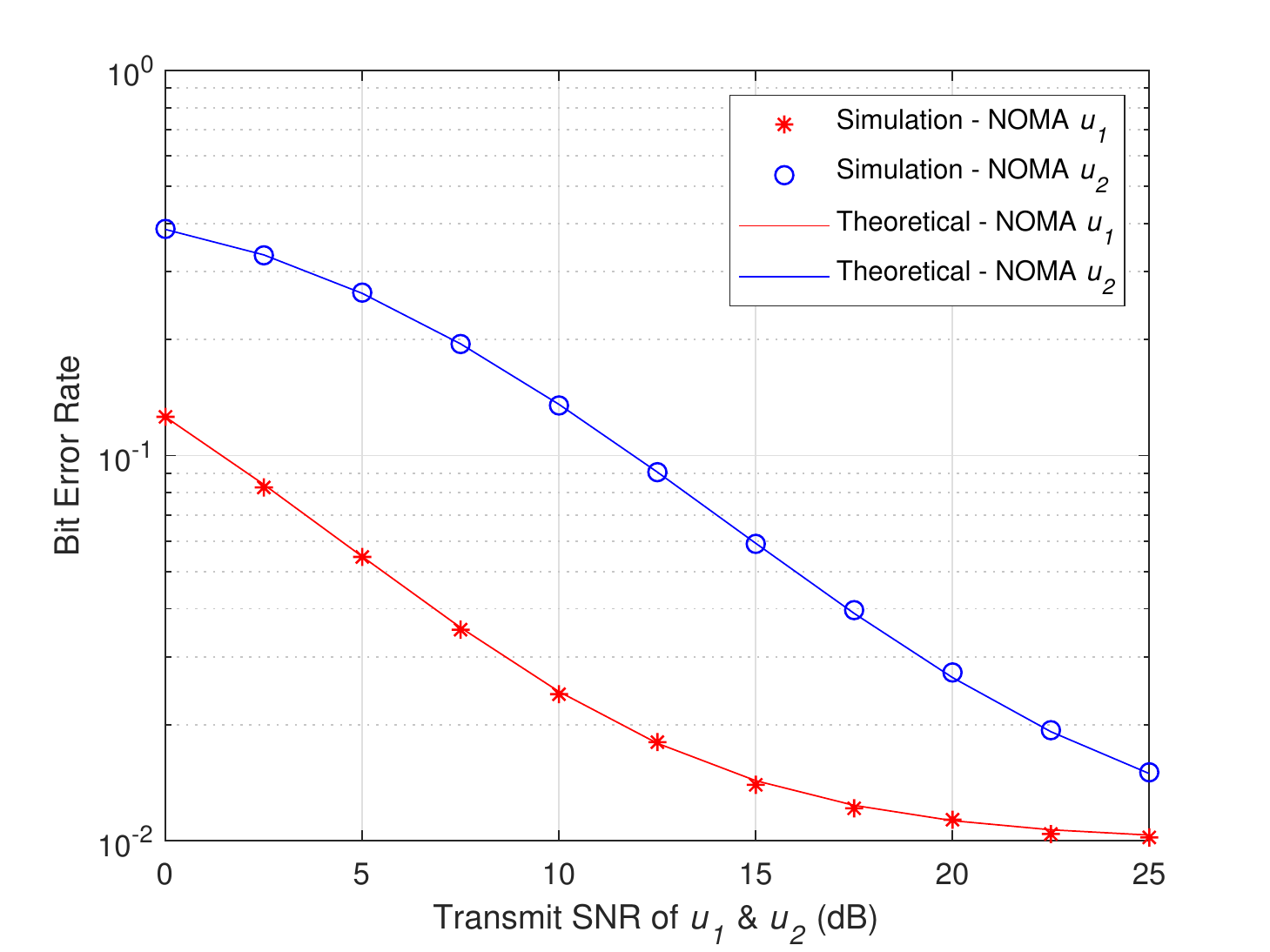}\label{fig:BER_plots_3}}
  \subfloat[$\Gamma_1=1, \Gamma_2=0.7, 0.2$]{\includegraphics[width= 3.35in]{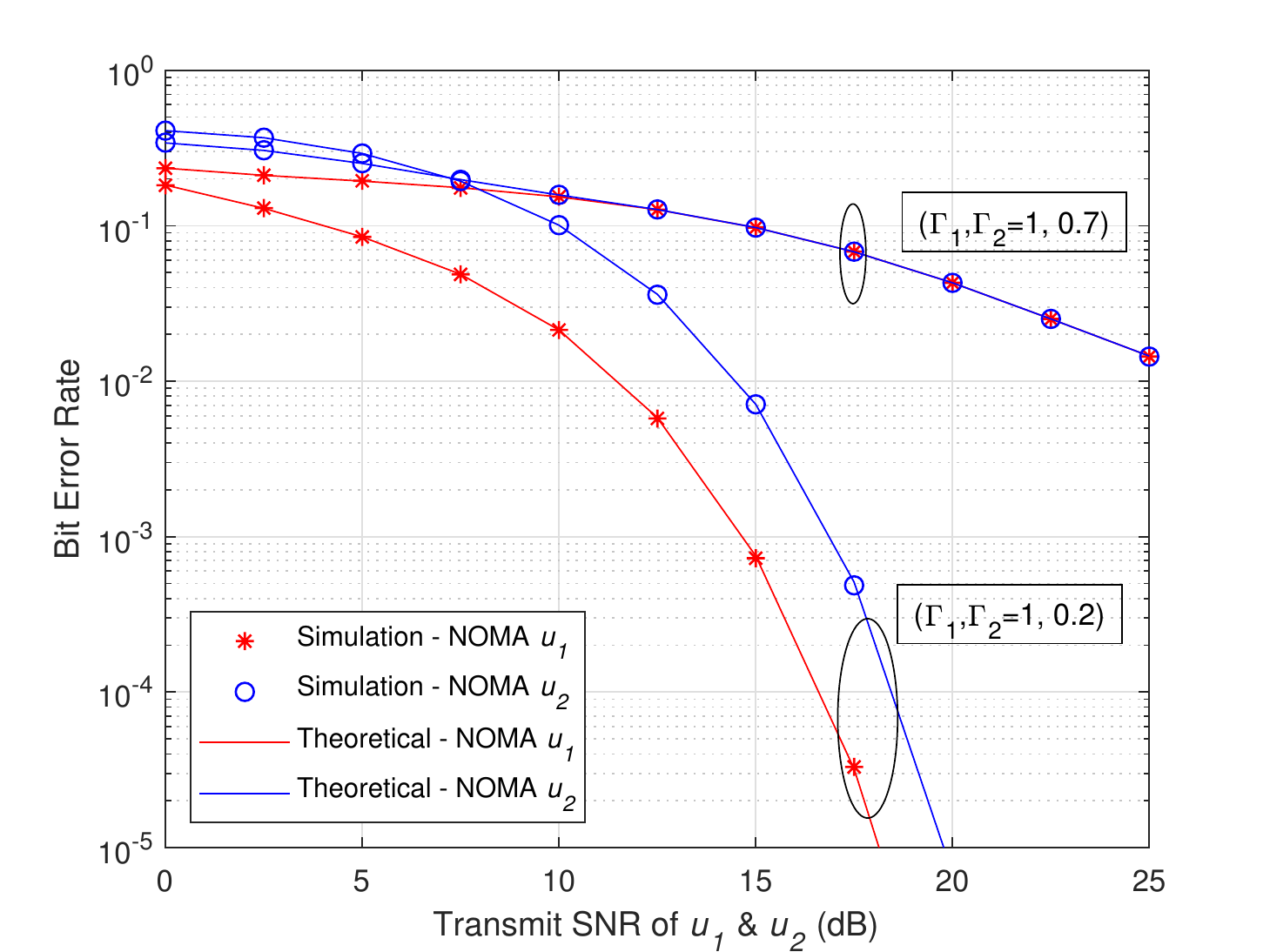}\label{fig:BER_plots_4}}
  \caption{BER plots of BSN-1 and BSN-2 for fading ($m_1=4, \Omega_1 =1,m_2=1, \Omega_2 =0$) and fading-free scenarios.}
  \label{fig:BER_plots}
\end{figure*}

This section investigates the performance of NOMA enhanced bistatic BackCom system consisting of a single cluster of two-BSNs and presents the numerical results by evaluating the BER expressions derived in the previous section. The results are validated with Monte Carlo simulations and are found to match the derived expressions in this paper. The channel between BSNs and BSR is modeled as Nakagami-$m$ fading channel, and both BSN and BSR are assumed to be equipped with a single antenna. The transmitted symbols for both users are selected uniformly from a BPSK constellation. Unless stated otherwise, the figures are plotted for fading channel conditions given as $m_1=4, \Omega_1 =1,m_2=1, \Omega_2 =0.5.$

\subsection{Analysis Validation}

In Fig. \ref{fig:BER_plots}, numerical and simulated BER of the NOMA enhanced BackCom system is plotted against transmit SNR of both BSNs. A Nakagami-$m$ fading channel ($m_1=4, \Omega_1=1$) and a Rayleigh fading channel ($m_2=1, \Omega_2=1$) is taken in Fig. \ref{fig:BER_plots}\ref{sub@fig:BER_plots_1}, \ref{fig:BER_plots}\ref{sub@fig:BER_plots_2} and \ref{fig:BER_plots}\ref{sub@fig:BER_plots_3} for the fading scenario of BSN-1 and BSN-2, respectively. Three different pairs of reflection coefficient values (\(\Gamma_1\), \(\Gamma_2\)) are considered for the analysis. As can be seen from the figure, numerical results obtained using (\ref{eq:Pu1_avergae}) and (\ref{eq:Pu2_average}) match the simulation results for different pairs of \(\Gamma_1\) and \(\Gamma_2\) values thus validating the theoretical analysis. It can also be observed that a larger separation in reflection coefficient values results in better BER behavior for NOMA enhanced BackCom. This is because, by lowering \(\Gamma_2\), the IUI experienced by BSN-1 is decreased, resulting in better BER performance of BSN-1 due to the efficient utilization of the NOMA principle. As the BER performance of BSN-2 depends on the successful SIC operation of BSN-1, therefore, lowering the IUI indirectly influences the performance of BSN-2 as evident from its improved performance.

The BER plots are also given in Fig. \ref{fig:BER_plots}\ref{sub@fig:BER_plots_4} for a fading-free scenario by taking an arbitrarily large value of $m_1$ and $m_2$ to simulate a pure AWGN channel. The fading-free scenario provides a significant improvement in BER performance. Furthermore, similar behavior of larger separation resulting in better performance is also observed in fading-free scenarios.


\subsection{Effects of the reflection coefficients}

\begin{figure}[hb]
  \vspace{-.75em}
  \centering
  \includegraphics[width=.65\columnwidth]{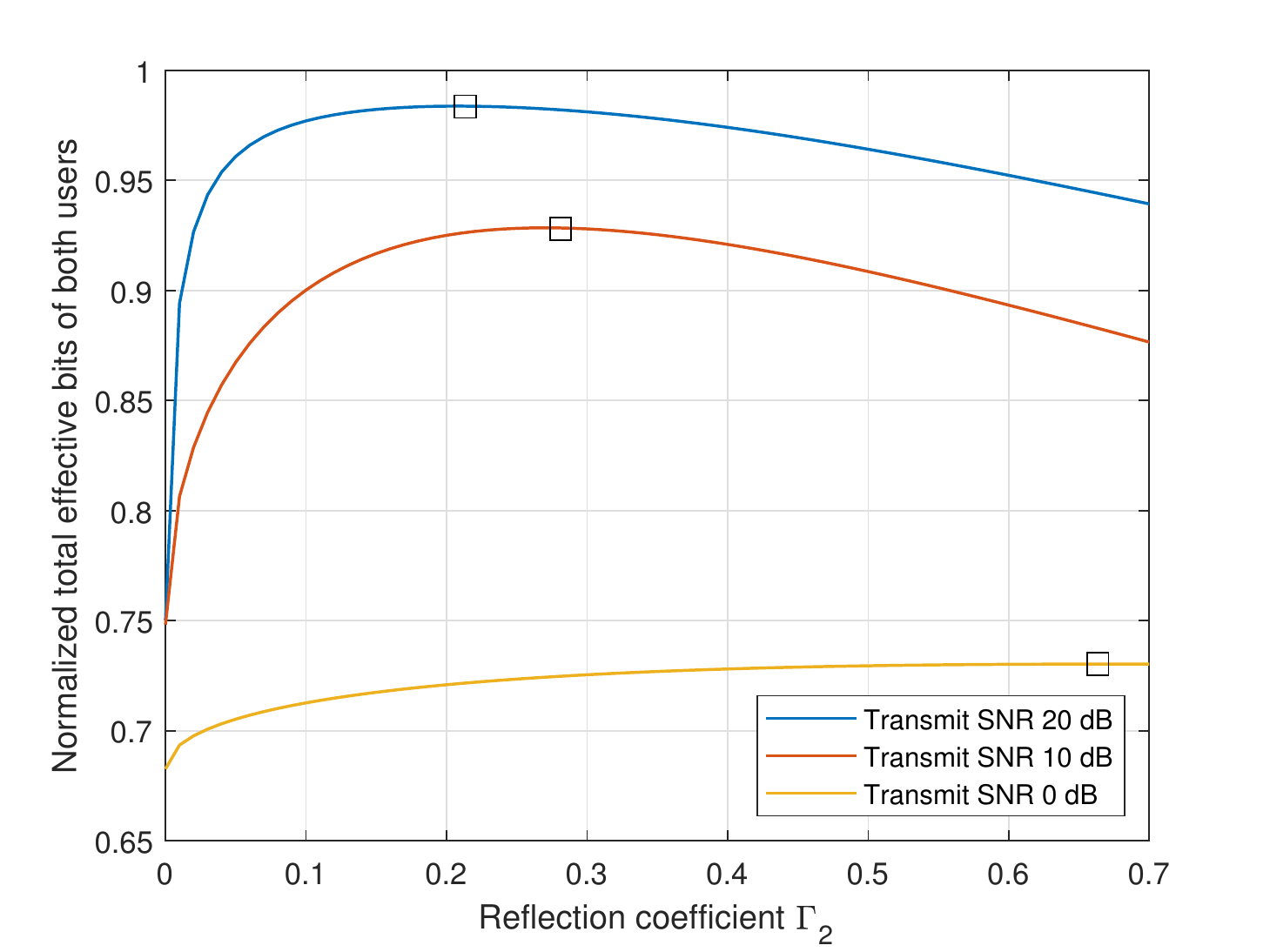}
  \caption{The normalized average of effectively decoded bits for BSN-1 and BSN-2 against $\Gamma_2$ while $\Gamma_1=0.7$.}
  \label{fig:Eff_bits_Plot_r2_vary}
\end{figure}

In Fig. \ref{fig:Eff_bits_Plot_r2_vary}, we investigate the effect of reflection coefficient on the normalized average of total effectively decoded bits of $u_1$ and $u_2$. The total effectively decoded bits correspond to the non-erroneous transmission of BSNs' bits over the total number of bits transmitted by BSN-1 and BSN-2. It is assumed that the reflection coefficient for the BSN-1, $\Gamma_1$, is set as 0.7 for the analysis. The results are plotted for three values of transmit SNRs, i.e., 0, 10, and 20 dB.

From Fig. \ref{fig:Eff_bits_Plot_r2_vary}, we can observe that the total normalized effective bits of $u_1$ and $u_2$ decrease with an increase in $\Gamma_2$ value. Therefore, the system performance is improved by setting a low value of $\Gamma_2$. This is again because, by setting a low value of $\Gamma_2$, the interference experienced by BSN-1 is reduced, thus, its BER performance is improved resulting in the greater non-erroneous transmission of bits by the system. However, it is to be noted from figure, that $\Gamma_2$ has a minimum value below which the performance of the system starts to degrade as BSN-2 is not able to decode itself at such a small value resulting in higher transmission errors. 

\begin{figure}[t!]
  \centering
  \subfloat[Fading with transmit SNR = 15 dB]{\includegraphics[width= 3.24in]{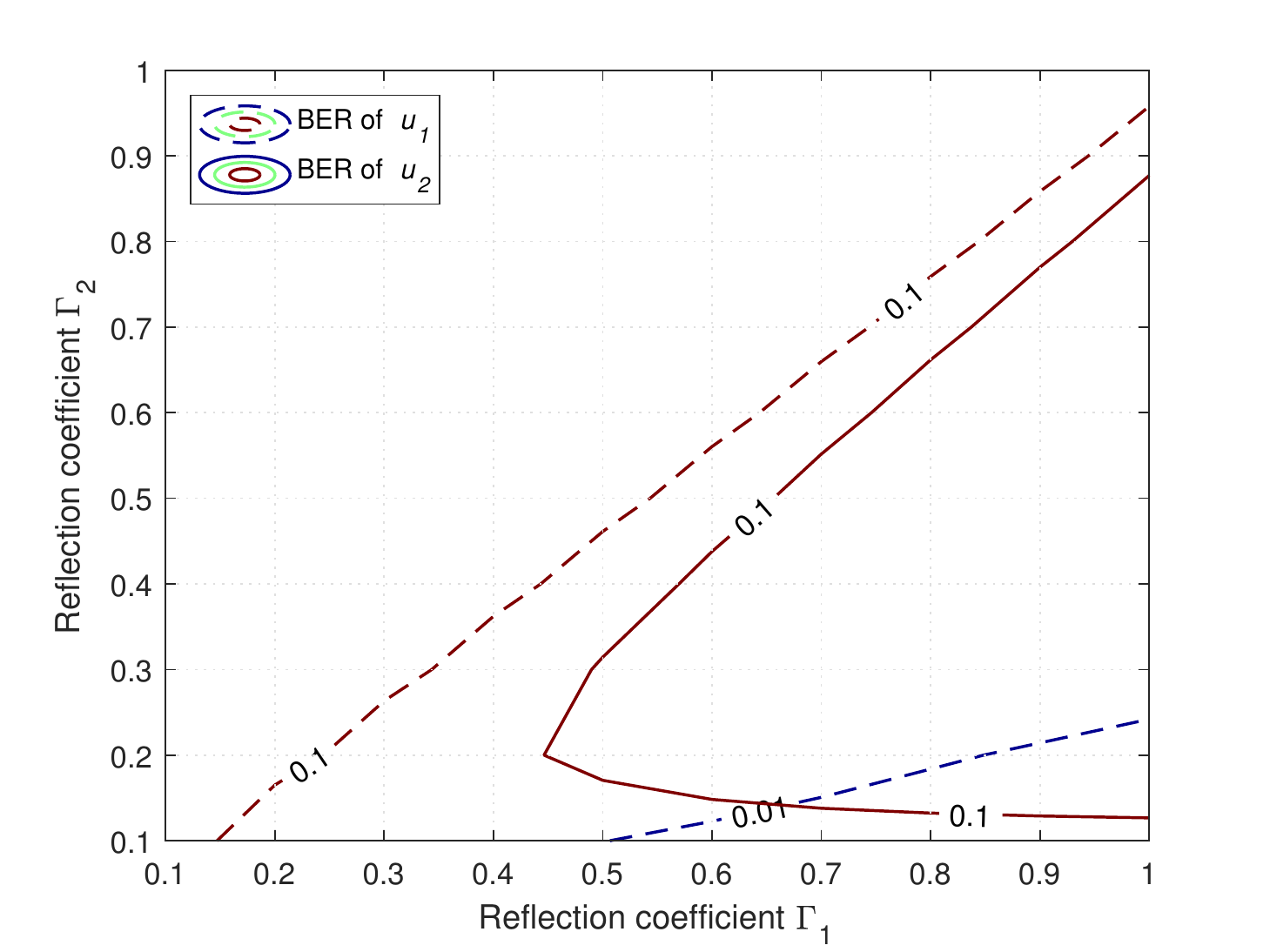}\label{fig:Contour_plots_1}}\\
  \subfloat[Fading with transmit SNR = 20 dB]{\includegraphics[width= 3.24in]{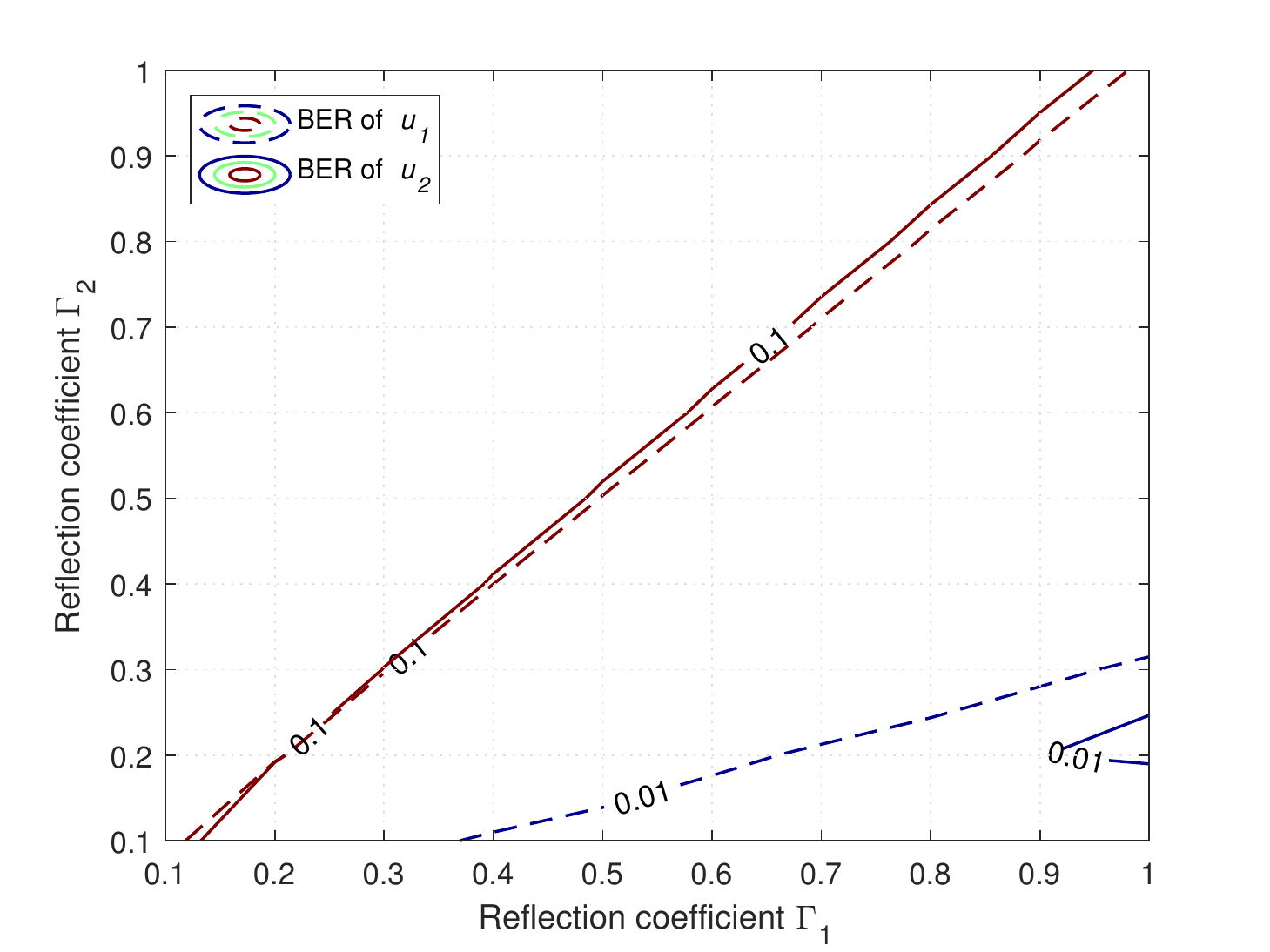}\label{fig:Contour_plots_2}}\\
  \subfloat[Fading-free with transmit SNR = 15 dB]{\includegraphics[width= 3.24in]{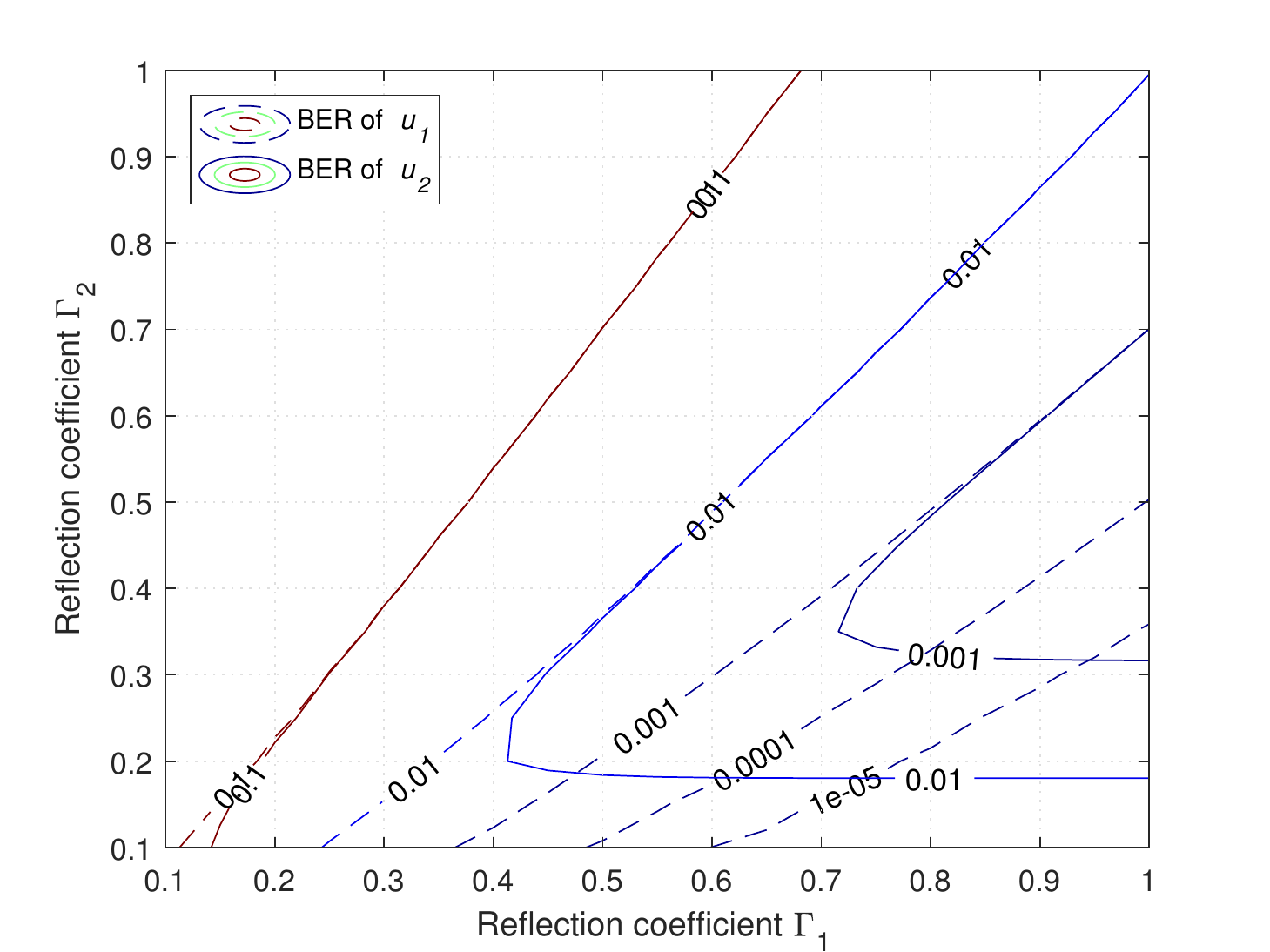}\label{fig:Contour_plots_3}}
  \caption{BER contour plot of BSN-1 and BSN-2 by varying $\Gamma_1$ and $\Gamma_2$ values for fading and fading-free scenarios}
  \label{fig:Contour_plots}
\end{figure}

The $\Gamma_2$ values are also marked for each transmit SNR which provides the best system performance in terms of the effective bits transmitted. For transmit SNRs of 0, 10, and 20 dBs, the optimal $\Gamma_2$ is 0.67, 0.27 and 0.21, respectively. It is evident from the fact that by lowering the transmit SNR, a greater $\Gamma_2$ value is needed by BSN-2 to decode itself successfully. We also find that the maximum normalized total effective bits turns out to be 0.9791 for the 20 dB SNR scenario. 

In Fig. \ref{fig:Contour_plots}, a contour plot of BER is plotted by varying the reflection coefficient pair of BSN-1 and BSN-2. The reflection coefficients of both BSNs are varied from 0.01 to a maximum value of 1 by always keeping \(\Gamma_1 > \Gamma_2\) as found from the previous result in Fig. \ref{fig:BER_plots}. The BER of BSN-1 is represented by a dotted line while the BER of BSN-2 is represented by a solid line. It can be observed from the contour plot that for any specific value of \(\Gamma_1\), there exists a range of \(\Gamma_2\) values smaller than \(\Gamma_1\) for which we can achieve acceptable performance in a NOMA-BackCom system. For the fading case (see Fig. \ref{fig:Contour_plots}\ref{sub@fig:Contour_plots_1} and \ref{fig:Contour_plots}\ref{sub@fig:Contour_plots_2}), it can be observed that very little improvement is possible in BER unless either the transmit SNR is increased by placing the BSNs closer to CE or fading channel condition is boosted by removing obstructions in line-of-sight (LoS) path. The contour plots of the fading-free case is plotted in Fig. \ref{fig:Contour_plots}\ref{sub@fig:Contour_plots_3} for improvement comparison due to less channel severity. 

\subsection{Effect of the Nakagami-$m$ fading parameter}

\begin{figure}[h]
  \centering
  \includegraphics[width=.7\columnwidth]{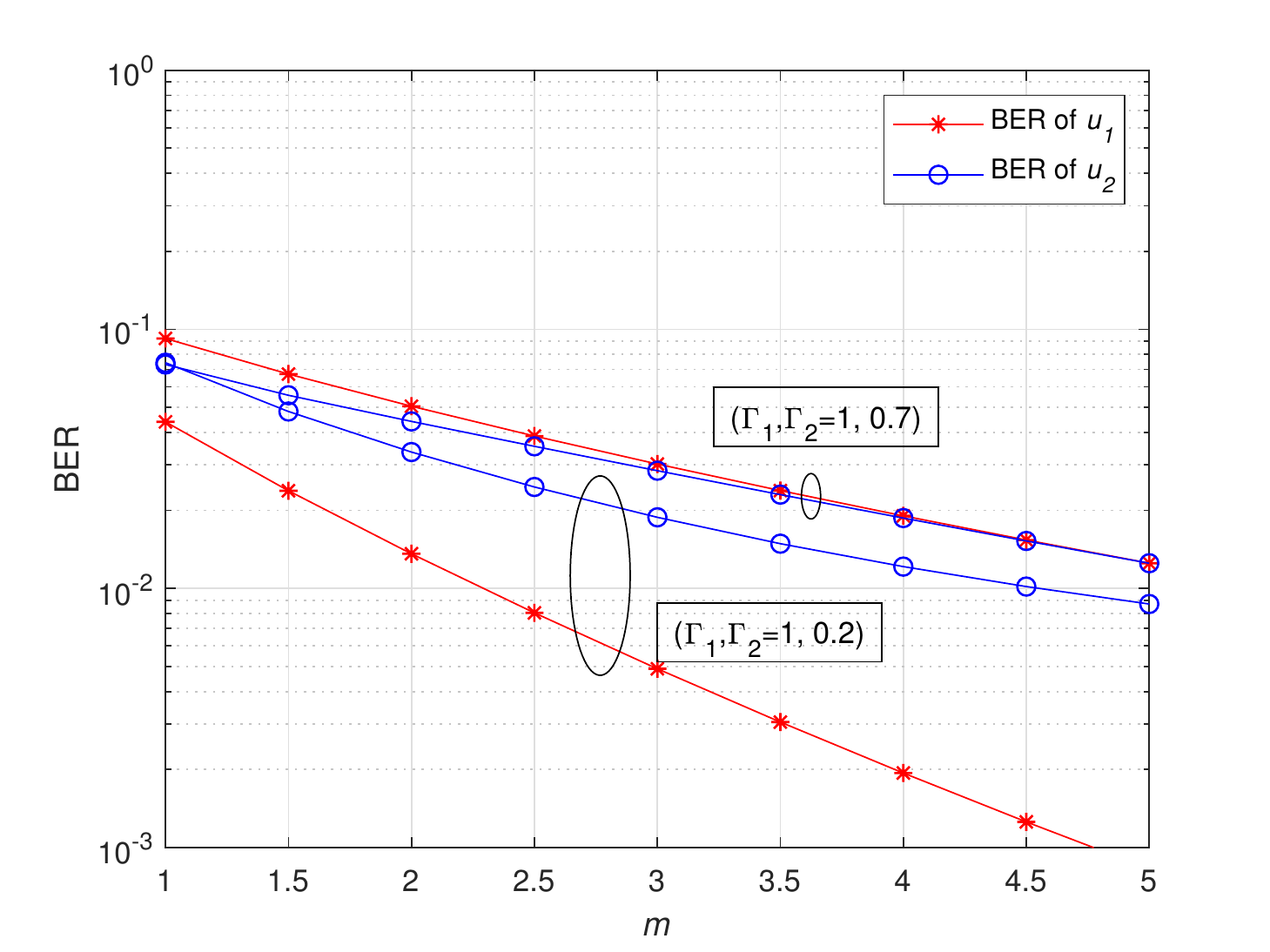}
  \caption{BER plots of BSN-1 and BSN-2 for various $m$ values.}
  \label{fig:m_vary}
\end{figure}

Now, we investigate the impact of the fading parameter $m$ on the BER performance of each BSN for a NOMA-BackCom system with two different sets of reflection coefficients. The BER plots are shown in Fig. \ref{fig:m_vary} for transmit SNRs of $u_1$ and $u_2$ as 20 and 15 dBs, respectively. As can be observed from figure, the BER performance of both users depend strongly on the fading parameters of the channel. It can be seen that for suitable reflection coefficient pairs, the fading parameters affects the performance of BSN-1 more than BSN-2.

\begin{figure}[t!]
  \centering
  \subfloat[Comparison of total effective bits]{\includegraphics[width= 4.2in]{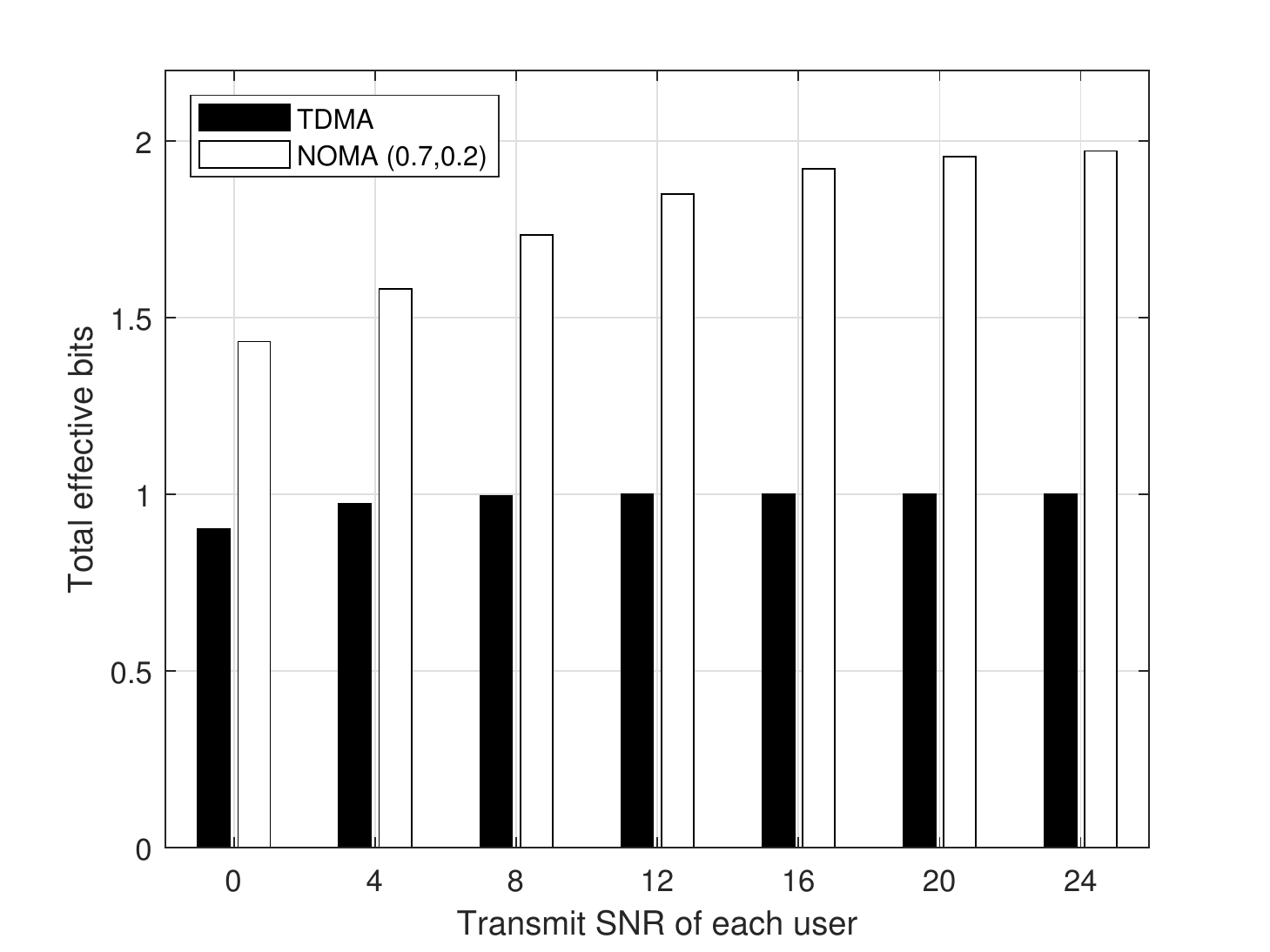}\label{fig:OMA-comparison_1}}
  \\
  \subfloat[Comparison of individual effective bits]{\includegraphics[width= 4.2in]{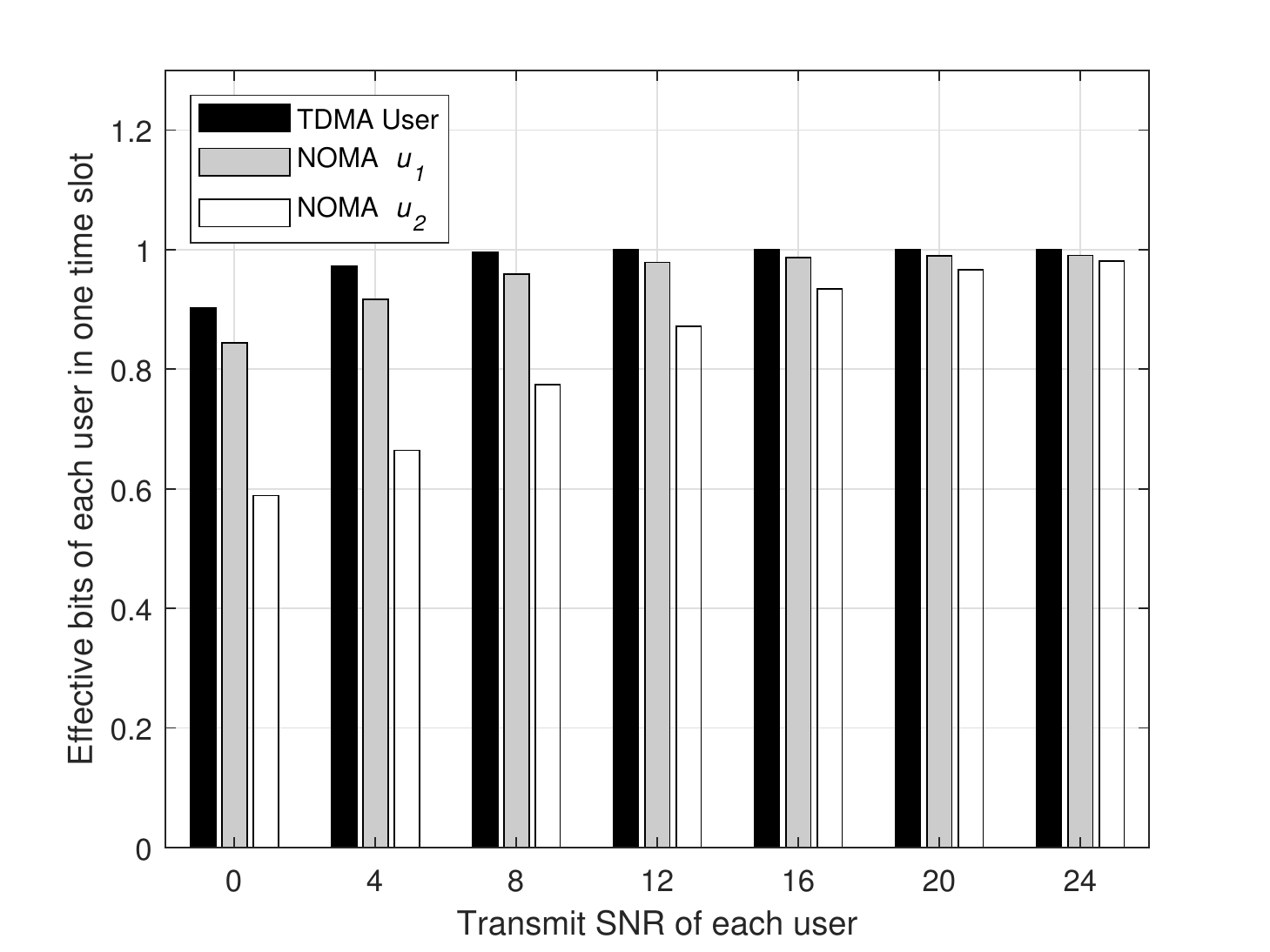}\label{fig:OMA-comparison_2}}
  \caption{Comparison of normalized effective bits transmitted of a NOMA-BackCom system ($\Gamma_1=0.7, \Gamma_2=0.2$) against OMA-TDMA scheme.}
  \label{fig:OMA-comparison}
  \vspace{-.75em}
\end{figure}

\subsection{Comparison with OMA-TDMA scheme}
In this section, we compare the performance of the NOMA-BackCom system with an OMA-TDMA transmission scheme in a BackCom system. Fig. \ref{fig:OMA-comparison} illustrates the increase in effective non-erroneous bits transmission by employing the NOMA scheme in a BackCom system of a single cluster with two BSNs as compared to an OMA-TDMA transmission scheme with two BSNs. It can be observed that the NOMA scheme indeed outperforms the OMA scheme due to the simultaneous transmission of two bits to the reader in a single time slot even though the OMA-TDMA scheme experiences no IUI from the second BSN. Individually, the TDMA BSN always has better BER performance as compared to NOMA BSN, i.e., a TDMA user has a greater number of successful transmissions as shown in Fig. \ref{fig:OMA-comparison} \ref{sub@fig:OMA-comparison_2}. However, the combined performance of NOMA scheme is better as the time spent by BSN on each time slot is doubled under NOMA scheme as compared to TDMA scheme. At high SNR though, the BER difference between an individual NOMA and TDMA user becomes negligible.

\section{Conclusion}
This work has presented the design and analysis of a NOMA enhanced bistatic BackCom system for a battery-less smart communication paradigm employed in an IoT scenario. We have derived the closed-form BER expressions for a cluster of two BSNs with imperfect SIC under Nakagami-$m$ fading channel. Furthermore, the PDFs of the sum and difference of two i.n.i.d Nakagami-$m$ distributions are also accurately approximated. All the derived expressions are verified with the simulations under different scenarios. Based on these expressions, we have evaluated the performance of the system in terms of the  reflection coefficients. We have also found that the increment of SNR with unsuitable reflection coefficients does not lead to a better system performance, hence highlighting the significance of setting proper reflection coefficients according to the scenario. This necessitates an optimization study as a future work where reflection coefficients can be optimized such that the system performance in terms of BER or effective transmitted bits can be improved. Further future extensions include BER analysis for higher modulation schemes with a higher number of BSNs in different fading environments.

\appendices
\section{K-S Test for the sum of two i.n.i.d  Nakagami-$m_i$ RVs}
The random number generation routine is repeated $N$ times to collect samples of the RV $Y$, i.e., $\{y_1,y_2,...,y_N \}$, with empirical cumulative distribution function (CDF) $\widehat{F}_Y $. The hypothesized CDF is that of Nakagami-$m_{R_1}$ distribution, $F_{R_1}$. The statistic used for goodness-of-fit known as K-S statistic is the maximum difference between the empirical CDF and hypothesized CDF, given by \cite{Press2002}
\begin{equation}\label{eq:KS_statistic}
  \widehat{D}_f =  \underset{{y}}{\operatorname{sup}}|\widehat{F}_Y(y_i)-F_{R_1}(y_i)|.
\end{equation}
The critical value is found to be $\widehat{c} = 0.0192$ for $N = 5000$ samples against the level of significance $\widehat{a} = 0.05$. The null hypothesis for testing is given as
\begin{equation}\label{eq:Null_hypothesis}
  H_0 : F_Y = F_{R_1}.
\end{equation}
The null hypothesis is accepted if $\widehat{D}_f \leq \widehat{c}$, i.e., $F_Y=F_{R_1}$ and rejected otherwise. The K-S test is conducted for three set of parameters, i.e., \{$m_1=1, \Omega_1 =1,m_2=1, \Omega_2 =1$\}, \{$m_1=3, \Omega_1 =1,m_2=1, \Omega_2 =1$\} and \{$m_1=3, \Omega_1 =1,m_2=3, \Omega_2 =0.5$\} and K-S statistic, $\widehat{D}_f $, calculated using (\ref{eq:KS_statistic}) are
\begin{equation}
  \begin{split}
    \widehat{D}_f &= 0.0122 \text{ for } m_1=1, \Omega_1 =1,m_2=1, \Omega_2 =1,\\
    \widehat{D}_f &= 0.0082 \text{ for } m_1=2, \Omega_1 =1,m_2=1, \Omega_2 =1,\\
    \widehat{D}_f &= 0.0077 \text{ for } m_1=3, \Omega_1 =1,m_2=3, \Omega_2 =0.5.\\
  \end{split}
\end{equation}
It can be observed, that $\widehat{D}_f \leq \widehat{c}$ for each parameter set and $H_0$ cannot be rejected. Hence, Nakagami-$m$ distribution closely matches the distribution of the sum of two Nakagami-$m$ RVs and can be used to approximate RV $Y$.

\section{K-S Test for the difference of two i.n.i.d  Nakagami-$m_i$ RVs}
Again, the Random number generation routine is repeated $N$ times to collect samples of the RV $Z$ , i.e., $\{z_1,z_2,...,z_N \}$, with empirical cumulative distribution function (CDF) $\widehat{F}_Z $. The hypothesized CDF is that of normal distribution, $F_{W_1}$.

The critical value is found to be $\widehat{c} = 0.0192$ for $N = 5000$ samples against the level of significance $\widehat{a} = 0.05$. The null hypothesis for testing is given as

\begin{equation}\label{eq:Null_hypothesis2}
  H_0 : F_Z = F_{W_1}.
\end{equation}

The K-S test is conducted for three set of parameters, i.e., \{$m_1=1, \Omega_1 =1,m_2=1, \Omega_2 =1$\}, \{$m_1=3, \Omega_1 =1,m_2=1, \Omega_2 =1$\} and \{$m_1=3, \Omega_1 =1,m_2=3, \Omega_2 =0.5$\} and K-S statistic, $\widehat{D}_f $, calculated using (\ref{eq:KS_statistic}) for each set, given as

\begin{equation}
  \begin{split}
    \widehat{D}_f &= 0.0109 \text{ for } m_1=1, \Omega_1 =1,m_2=1, \Omega_2 =1,\\
    \widehat{D}_f &= 0.0141 \text{ for } m_1=2, \Omega_1 =1,m_2=1, \Omega_2 =1,\\
    \widehat{D}_f &= 0.0140 \text{ for } m_1=3, \Omega_1 =1,m_2=3, \Omega_2 =0.5.\\
  \end{split}
\end{equation}

$H_0$ cannot be rejected for each parameter set as $\widehat{D}_f \leq \widehat{c}$. Hence, normal distribution closely matches the distribution of the difference of two Nakagami-$m$ RV and can be used to approximate RV $Z$.

\section{Proof of Lemma 2}
Let $W$ be a normal distributed RV with parameters mean $\mu$ and variance $\sigma^2$, i.e., $W\sim N(\mu,\sigma^2)$ and $Z$ be a standard normal distributed RV, i.e., $Z\sim N(0,1)$. Both being independent, then $\mathcal{Q}(w)=\mathbb{P}(Z>w)$ as $\mathcal{Q}(x)$ is the probability that a standard normal RV takes a value greater than $x$. This is similar to $\Lambda$ because the integrand in $\Lambda$ is the product of $\mathbb{P}(Z>w) =\mathbb{P}(Z>W|W=w)$ and the marginal density of normal RV, $W$, i.e., $f_W(w)$, which comes out to be $\mathbb{P}(Z>W)$. As we know that distribution of the difference of two normal RVs is again a normal RV with subtracted means but added variances, i.e., $Z-W \sim N(-\mu,\sigma^2+1)$, therefore
\begin{equation}
  \Lambda = \mathbb{P}(Z>W) = \mathbb{P}(Z-W>0) = \mathcal{Q}\bigg(\frac{\mu}{\sigma^2+1}\bigg).
\end{equation}


\ifCLASSOPTIONcaptionsoff
  \newpage
\fi

\bibliographystyle{IEEEtran}
\bibliography{IEEEabrv,TII_References} 




\end{document}